\documentclass[aps,pr,preprint,amsmath,amssymb,showkey]{revtex4-2}
\usepackage{graphicx}
\usepackage{float} 
\usepackage{newtxtext}
\usepackage{newtxmath}

\usepackage{dcolumn}
\usepackage{bm}
\usepackage[utf8]{inputenc}
\usepackage[T1]{fontenc}
\usepackage{xcolor}

\usepackage{hyperref}
\usepackage{cleveref}
\hypersetup{colorlinks = true, linkcolor = blue,  urlcolor = blue, citecolor = blue} 

\begin{document}

\title{Onset instability of inverted flags clamped by a cylinder}

\author{Haokui Jiang{\color{blue}$^1$}}
\author{Yujia Zhao{\color{blue}$^1$}}
\author{Burigede Liu{\color{blue}$^2$}}
\author{Shunxiang Cao{\color{blue}$^{1,}$}}%
\email{caoshunxiang@sz.tsinghua.edu.cn}
 \affiliation{{\color{blue}$^1$}Institute for Ocean Engineering, Shenzhen International Graduate School, Tsinghua University, Shenzhen 518055, China}
 
\affiliation{{\color{blue}$^2$}Department of Engineering, University of Cambridge, Trumpington Street, Cambridge, CB2 1PZ, UK}

\begin{abstract}
We numerically investigate the hydrodynamic characteristics and analyze the instability mechanism of a two-dimensional inverted flag clamped by a cylinder. Two transition routes and a total of six kinds of solutions exist under this configuration for different diameters of cylinders due to complex bifurcations. Specifically, for small cylinders, the undeformed equilibrium transitions to static deformed equilibrium through a supercritical pitchfork bifurcation, which is judged by the weakly nonlinear analysis together with the global linear instability analysis. The instability mechanism is the lifting effect of the steady structure mode working at the leading edge of the elastic plate. For large cylinders, another unstable fluid mode (decoupled with structure mode) causes the disappearance of the static undeformed and deformed equilibrium, replaced by a small amplitude flapping. The structure mode and the flow mode mainly contribute to the growth of perturbations in plate and downstream cylinder regions respectively, which can excite multi-mode oscillating transition analyzed by proper orthogonal decomposition. Moreover, we find there is a critical diameter $D_c$ dividing the pitchfork bifurcation and Hopf bifurcation, and $D_c$ decreases with the increase of Reynolds number. Finally, we prove downstream vortex shedding can induce upward vortex-induced vibration of the plate and further improve the efficiency of energy transfer from the fluid to the structure during small-deflection flapping. 

\end{abstract}

\maketitle


\section{Introduction}
\label{sec:introduction}
Exploring the onset instability of flow past inverted flags can improve our understanding of the flapping of biological structures, such as plant leaves in wind \cite{zhang2022effect,dong2024flag} and has potential application in engineering, for example, the small-scale energy harvesting \cite{orrego2017harvesting,silva2019simultaneous}. This issue has received widespread attention in recent years, and researchers have approached it by experiments, numerical simulation, and analytical modeling \cite{umair2022experimental, lim2022numerical, tavallaeinejad2020instability}. It is notable that most numerical and analytical studies on this topic concern the simplest single elastic plate configuration and neglect the influence of clamps. However, clamps with thickness far larger than the elastic plate are ubiquitous for practical inverted flags system \cite{shen2024coupled, cioncolini2019experimental,hu2020passive}. The fundamental fluid mechanics in this fluid-structure interaction (FSI) problem have not been clearly revealed, potentially inhibiting a more extensive application of inverted flags. Therefore, we simplify clamps as a cylinder and systematically study the hydrodynamics and instability of inverted flags clamped by a cylinder at moderate Reynolds numbers here. In the following, we summarize the work of hydrodynamics and bifurcations of inverted flags and concisely discuss the position of the current work in the literature.

~\\
\centerline{1.1 \emph{Hydrodynamic behaviors of inverted flags}}
~\

Kim \emph{et al}. explored dynamics of inverted flags experimentally by increasing Reynolds number $Re$ from $Re = 10^4$ to $10^5$, they identified three dynamical regimes, namely the stretched-straight mode, the flapping mode and the fully deflected mode \cite{kim2013flapping}. In this range of Reynolds number, Sader \emph{et al}. concluded that the flapping mode of inverted flags was a vortex-induced vibration (VIV) by experimental measurement and scaling analysis \cite{sader2016large}. Fan \emph{et al}. studied the effects of the shape of inverted flags on the critical Reynolds numbers of onset and cessation of flapping \cite{fan2019effect}. Yu \emph{et al}. conducted the relationship of oscillating amplitude with the Reynolds numbers for these three dynamical regimes by experiments in the $Re = 4660-7600$ flow regime \cite{yu2017vortex}. These experimental works concentrated on hydrodynamics performance of inverted flags in a high Reynolds numbers regime, while more behaviors were found by numerical investigations at moderate Reynolds number ranges ($Re \leq 1000$). 
 
Ryu \emph{et al}. \cite{ryu2015flapping} investigated the flapping dynamics of two-dimensional (2D) inverted flags at Reynolds numbers $Re \le 250$ by using the immersed boundary method (IBM) \cite{cao2021spatially}. They found the stretched-straight mode and the flapping mode vanish when $Re < 50$ and studied the transition routes about the bending stiffness $K_B$ for $100 < Re < 250$. Gurugubelli and Jaiman performed inverted flags no longer exhibit the periodic flapping due to the absence of vortex shedding; instead, it undergoes a large static deformation in the low-$Re$ regime $Re\in [0.1, 50]$ by using an arbitrary Lagrangian-Eulerian (ALE) method \cite{gurugubelli2019large}.  Tang \emph{et al}. found inverted flags undergo flapping motion about the deflected shape called deflected-flapping mode when increasing $Re$ or decreasing the bending parameter by using IBM in a range of Reynolds number $100 - 500$ \cite{tang2015dynamics}. Shoele and Mittal illustrated that the inverted flags were dynamically insensitive to the mass ratio $M$ (low- or high-amplitude symmetric flapping phase) for Reynolds number change from 25 to 800 \cite{shoele2016energy}. Goza \emph{et al}. showed that chaotic flapping could occur at a moderate Reynolds number of 200, which had been observed experimentally for Reynolds number of $O(10^4)$. Moreover, large-amplitude flapping arises even for sufficiently heavy flags when $Re < 50$ \cite{goza2018global}. 

~\\
\centerline{1.2 \emph{Bifurcations and instability mechanism of inverted flags}}
~\

Inverted flags has a wide range of dynamical regimes due to complex bifurcations, such as supercritical, subcritical or saddle-node. Rinaldi and Païdoussis found a weak flutter-like instability when steady undeformed inverted flags bifurcated to a large static deformation at low flow velocities, and it changed to divergence instability at higher flow velocities \cite{rinaldi2012theory}. Thereafter, their team proved these hydrodynamics qualitatively by using a inverted cylinder in confined axial flow \cite{abdelbaki2018nonlinear}. Sader \emph{et al}. also found a divergence instability in their experiments. Moreover, they observed inverted flags undergo a saddle-node bifurcation, that is zero-deflection equilibrium transitions to large-amplitude deflected equilibrium abruptly  \cite{sader2016stability}. Tavallaeinejad \emph{et al}. found that a low aspect ratio (i.e. height to length ratio) flag undergoes a static divergence via a subcritical pitchfork bifurcation followed by a saddle-node bifurcation. However, beyond a critical aspect ratio, the subcritical bifurcation is replaced by a supercritical bifurcation \cite{tavallaeinejad2018nonlinear}. Goza \emph{et al}. conducted a global linear instability and proved the static deformed-equilibrium transition to small-deflection flapping by a supercritical Hopf bifurcation. Moreover, they found a non-classical VIV mechanism for large-amplitude flapping with heavier flags \cite{goza2018global}. 

Tang \emph{et al}. found the bifurcation from stretched-straight phase to flapping phase was bistability, as well as flapping phase transition to deflected phase by a subcritical bifurcation, which performed as a hysteresis \cite{tang2015dynamics}. Instability mechanism of flapping of the light flag is attributed to vortex-induced vibration and lock-in phenomenon, i.e. vortex shedding and flapping occur in synchrony, with flag motion and fluid forces sharing the same dominant frequency  \cite{shoele2016energy}. Sader \emph{et al}. argued that large amplitude flapping motion of inverted flag was also a vortex-induced vibration \cite{khalak1999motions,sarpkaya2004critical}. Recently, Tavallaeinejad \emph{et al}. found that inverted flags were flapping about a deflected equilibrium (performing as an asymmetric flapping or deflected-flapping mode) via two saddle-node bifurcations. Continuing to enhance Reynolds number, inverted flags entered a highly deflected state on one side by a Hopf bifurcation \cite{tavallaeinejad2020instability, tavallaeinejad2020nonlinear}.

~\\
\centerline{1.3 \emph{The position and structure of the current work}}
~\

The present work motivates us to investigate hydrodynamics, bifurcations and instability mechanism of inverted flags clamped by a cylinder at moderate Reynolds numbers. We care about the effect of the  cylinder's diameter on transition, so we fix Reynolds number and mass ratio and change the flag’s stiffness and diameter in most direct numerical simulations (DNS). Then, bifurcations and instability mechanisms are explained by global linear and weakly nonlinear analysis and proper orthogonal decomposition (POD) analysis \cite{sirovich1987turbulence}. Finally, with the instability and energy transfer mechanisms being illustrated clearly, they will expand practical applications of inverted flags.

The manuscript is organized as follows: In Sec. {\color{blue}II}, we outline the physics problem under investigation, including the governing equations, global linear instability analysis, and proper orthogonal decomposition. The numerical results such as linear eigenmodes, nonlinear hydrodynamics, transition and energy transfer are then presented and discussed in Sec. {\color{blue}III}. Concluding remarks are provided in Sec. {\color{blue}IV}.

\section{PROBLEM FORMULATION}
\label{sec:formulationa}
\subsection{Problem definition}
\label{subsec:DNS}

\begin{figure}[H]
\centerline{\includegraphics[width=0.8\linewidth]{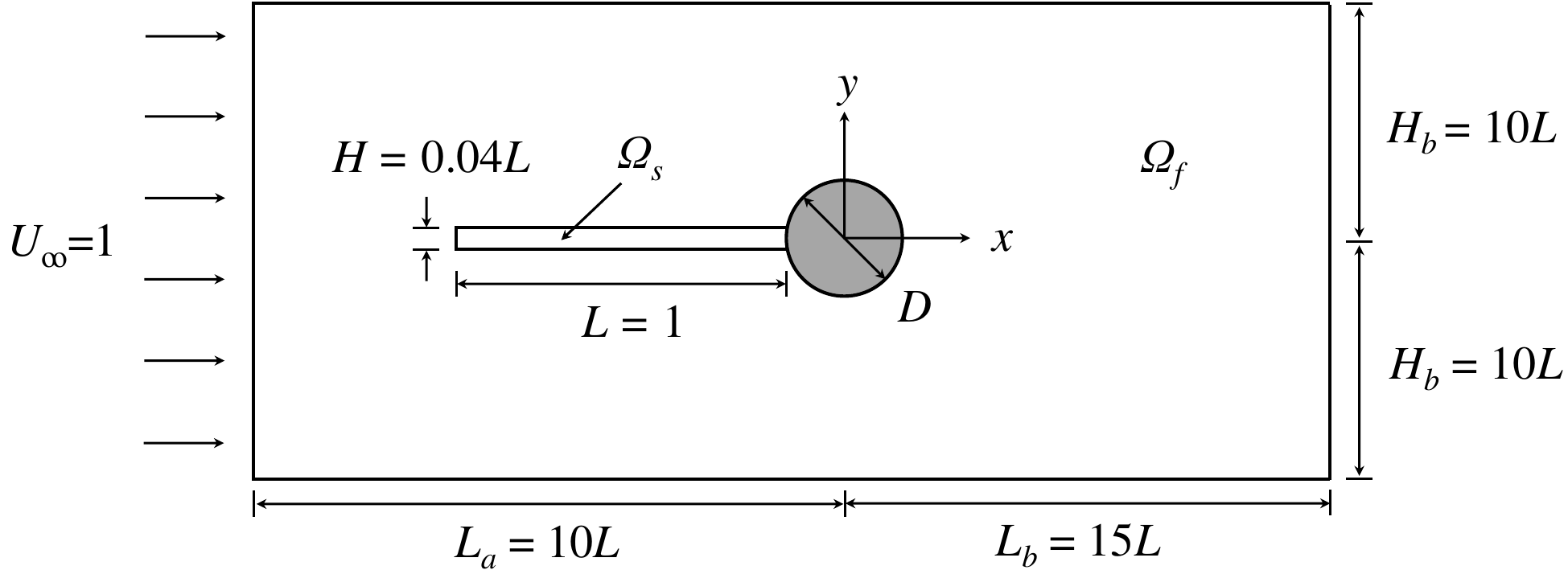}}
\caption{Sketch of the elastic plate clamped on the rigid cylinder and immersed in a uniform incoming flow field. Computational domain of size $[25L \times 20L]$. }
\label{fig:Fig1}
\end{figure}

We consider a two-dimensional thin elastic plate $\Omega_{s}$ with a length $L$ and thickness $H$ interacting with an incompressible uniform axial flow $\Omega_{f}$. As shown in figure \ref{fig:Fig1}, the plate is clamped on the left side of a rigid circular cylinder of diameter $D$. A Cartesian coordinate system with its origin at the center of the cylinder is used. The $x$-coordinate aligns with the streamwise direction, whereas the $y$-coordinate corresponds to the lateral direction.

The governing equation of this fluid-structure interaction problem includes fluid sub-problem, solid sub-problem, extension sub-problem and boundary conditions. First, the Navier-Stokes equations governing an incompressible flow in an arbitrary Lagrangian-Eulerian reference frame are 
\begin{equation}
\bm{\nabla} \cdot \left[ \bm{\Phi} \left( {{\eta }_{e}} \right)\bm{u} \right]=0  \quad     \text{in} \quad {{\Omega }_{f}},
\label{eq:NS1}
\end{equation}
\begin{equation}
J\left( {{\eta }_{e}} \right)\frac{\partial \bm{u}}{\partial t}+\left[ \left( \bm{\nabla}\bm{ u} \right)\bm{\Phi} \left( {{\eta }_{e}} \right) \right]\left( \bm{u}-\frac{\partial {{\eta }_{e}}}{\partial t} \right)-\bm{\nabla} \cdot \bm{\Sigma} \left( \bm{u},p,{{\eta }_{e}} \right)=0  \quad     \text{in} \quad {{\Omega }_{f}},
\label{eq:NS2}
\end{equation}
where ${{\eta }_{e}}$ is extension displacement field used to describe the deformation of the fluid domain in a Lagrangian framework; $\bm{u} = (u_x, u_y)$ is the velocity of the fluid; $\bm{\Phi} \left( {{\eta }_{e}} \right)=J\left( {{\eta }_{e}} \right){{\left( \bm{I}+\bm{\nabla} {{\eta }_{e}} \right)}^{-1}}$ is deformation operator, which expresses how an infinitesimal surface in the instantaneous configuration is transformed in the reference configuration \cite{aris2012vectors}; $\bm{I}$ is the unit matrix; $J\left( {{\eta }_{e}} \right)$is the transformation Jacobian. $\bm{\Sigma} $ is the first Piola-Kirchhoff stress tensor for the fluid, written as 
\begin{equation}
\bm{\Sigma} \left( \bm{u},p,{{\eta }_{e}} \right)=\left\{ -p\bm{I}+\frac{1}{{Re}}\frac{1}{J\left( {{\eta }_{e}} \right)}\left[ \left( \bm{\nabla} \bm{ u} \right)\bm{\Phi} \left( {{\eta }_{e}} \right)+{{\left( \bm{\nabla} \bm{u} \right)}^{T }}\bm{\Phi} {{\left( {{\eta }_{e}} \right)}^{T }} \right] \right\}\bm{\Phi} {{\left( {{\eta }_{e}} \right)}^{T }}.
\end{equation}

Then, the solid elasticity equation is defined as, 
\begin{equation}
{{M}_{s}}\frac{{{\partial }^{2}}{{\eta }_{s}}}{\partial {{t}^{2}}}-\bm{\nabla} \cdot \left[ \bm{F}\left( {{\eta }_{s}} \right)\bm{S}\left( {{\eta }_{s}} \right) \right]=0 \quad     \text{in} \quad {{\Omega }_{s}},
\label{eq:Cyl_motion1}
\end{equation}
where ${{M}_{s}}$ represents the mass of structure; $\eta_s$ is the solid displacement field; $\bm{F}\left( {{\eta }_{s}} \right)=\bm{I}+\bm{\nabla} {{\eta }_{s}}$ is the deformation gradient, $\bm{S}\left( {{\eta }_{s}} \right)={{\lambda }_{s}}tr\left[ \bm{E}\left( {{\eta }_{s}} \right) \right]I+2{{\mu }_{s}}\bm{E}\left( {{\eta }_{s}} \right)$ is the second Piola-Kirchhoff stress tensor written here for a compressible Saint-Venant-Kirchhoff material \cite{ogden1997non}; ${{\lambda }_{s}}$ and ${{\mu }_{s}}$ are the Lame coefficients; $\bm{E}\left( {{\eta }_{s}} \right)={1}/{2}\;\left[ \bm{F}{{\left( {{\eta }_{s}} \right)}^{T}}\bm{F}\left( {{\eta }_{s}} \right)-\bm{I} \right]$ is the Green-Lagrange strain tensor.
\begin{equation}
\bm{\nabla} \cdot {{\bm{\Sigma} }_{e}}\left( {{\eta }_{e}} \right)=0  \quad     \text{in} \quad {{\Omega }_{e}},
\end{equation}
\begin{equation}
{{\eta }_{e}}-{{\eta }_{s}}=0  \quad  \text{on}  \quad  \Gamma,
\end{equation}
where ${\bm{\Sigma} }_{e}$ is the extension operator; $\Gamma$ is the interface of fluid and structure. In order to ensure a smooth propagation of large deformation, biharmonic equation is used here \cite{helenbrook2003mesh}. 

Finally, the velocity and stress continuity at the fluid-structure interface give
\begin{equation}
\bm{u}-\frac{\partial {{\eta }_{e}}}{\partial t}=0  \quad \text{on}\quad \Gamma,
\end{equation}
\begin{equation}
\bm{\Sigma} \left( \bm{u},p,{{\eta }_{e}} \right)n-\left[ \bm{F}\left( {{\eta }_{s}} \right)\bm{S}\left( {{\eta }_{s}} \right) \right]n=0  \quad \text{on}\quad \Gamma.
\end{equation}

In addition to this non-dimensional coefficient, the fluid-elastic configuration is governed by three non-dimensional parameters, defined here with $L$ and $U$ as characteristic length and velocity. The flapping dynamics of the conventional plate is strongly influenced by three non-dimensional parameters, namely Reynolds number $Re$, mass ratio $M$ and bending stiffness $K_B$, which are defined as:
\begin{equation}
Re=\frac{{{\rho }_{f}}{{U}_{\infty }}L}{\mu },\text{  }M=\frac{{{\rho }_{s}}H}{{{\rho }_{f}}L},\text{  }{{K}_{B}}=\frac{B}{{{\rho }_{f}}U_{\infty }^{2}{{L}^{3}}},
\end{equation}
where $\rho_f(\rho_s)$ is the fluid (structure) density, ${{U}_{\infty }}=1$ is the free-stream velocity, $L=1 $ is the flag length $\mu$ is the viscosity of the fluid, $H$ is the flag thickness and $B$ represents the flexural rigidity defined as $B = EH^3/[12(1 - v^2)]$ (where $E$ is the Young’s modulus of the elasticity and $v=0.34$ is the Poisson’s ratio). In this work, we fix $H = 0.04$ and $\rho_f = \rho_s$, therefore $M = 0.04$.

\subsection{Global linear stability analysis of steady solution}
\label{subsec:RL}

Global linear instability analysis here is performed based on the steady non-linear fluid-structure equations states solved by the Newton method \cite{ghattas1995variational}. Steady states are time-independent solutions of governing equations {\color{blue}(1)-(8)}, $Q=\left( U,P,{{\Pi }_{s}},{{\Pi }_{e}} \right)$ for the velocity, pressure, solid displacement and extension displacement fields, respectively. The fluid loads exactly balance the elastic restoring force, so the fluid-structure interface does not move. The fluid velocity is equal to zero at this interface and the velocity of the fluid domain in Eq. {{\color{blue}(2)} also vanishes, i.e., ${\partial {{\Pi }_{e}}}/{\partial t}\;=0$. 

The steady-state solutions are computed in the stress-free configuration, that is, the spatial domain $\Omega_s$ occupied by the structure when no external stresses are applied on it, while the linear stability analysis is performed in the steady deformed configuration. The steady-state solutions expressed in the steady configuration are
\begin{equation}
(U,P)(\bm{x}) = (U_0,P_0)(\bm{x}_0), \quad \text{and} \quad (\Pi_s,\Pi_e)(\bm{x}) = (0,0),
\end{equation}
where $U_0$ and $P_0$ are solutions at the stress-free configuration coordinate system $\bm{x}_0$. The linear FSI problem is obtained by decomposing the flow variable as a sum of the steady state and perturbation, 
\begin{equation}
\begin{aligned}
   \left( \bm{u},p \right)\left(\bm{ x},t \right)=\left( U,P \right)\left( \bm{x} \right)+\left( \hat{\bm{u}},\hat{p} \right)\left( \bm{x} \right){{e}^{-i\omega t}}\quad&\text{ in}\quad{{\Omega }_{f}}, \\ 
  \left( {{\eta }_{s}},{{\eta }_{e}} \right)\left( \bm{x},t \right)=\left( 0,0 \right)+\left( {{{\hat{\eta }}}_{s}},{{{\hat{\eta }}}_{e}} \right)\left( \bm{x} \right){{e}^{-i\omega t}}\quad&\text{ in}\quad {{\Omega }_{f}}\cup{{\Omega }_{s}}, \\ 
\end{aligned}
\label{eq:Cyl_decomposed}
\end{equation}
where perturbation is decomposed in the form of global eigenmodes; $\left( \hat{\bm{u}},\hat{p},{{{\hat{\eta }}}_{s}},{{{\hat{\eta }}}_{e}} \right)$ is the shape function and the complex-valued $\omega = \omega_r + i\omega_i$ is the circular frequency of the perturbation. The real part $\omega_i$ indicates the growth ($\omega_i>0$) or decay of the mode, while the imaginary part $\omega_r$ gives its oscillation frequency. After substituting the above decompositions into the updated coordinate ALE formulation, from which the governing equations for the base states are subtracted and the first-order terms are retained, the linear system for the fluid-structure system reads as
\begin{equation}
\bm{\nabla} \cdot \hat{\bm{u}}+\bm{\nabla} \cdot \left[ \bm{\Phi }'\left( {{{\hat{\eta }}}_{e}} \right)U \right]=0  \quad \text{in}\quad{{\Omega }_{f}},
\label{eq:Linear1}
\end{equation}
\begin{equation}
\begin{aligned}
  \omega \left[ \hat{\bm{u}}-\left( \bm{\nabla} U \right){{{\hat{\eta }}}_{e}} \right]-\bm{\nabla} \cdot &\left[ \bm{\Sigma }'\left( {{{\hat{\eta }}}_{e}},U,P \right) +\bm{\sigma} \left( \hat{\bm{u}},\hat{p} \right) \right] \\ 
 &\qquad +\left( \bm{\nabla} U \right)\hat{\bm{u}}+\left( \bm{\nabla} \hat{u} \right)U+\left( \bm{\nabla} U \right){\Phi }'\left( {{{\hat{\eta }}}_{e}} \right)U=0 \quad \text{in}\quad{{\Omega }_{f}}, \\ 
\end{aligned}
\label{eq:Linear2}
\end{equation}
\begin{equation}
{{\omega }^{2}}\frac{{{M}_{s}}}{J\left( {{\Pi }_{s}} \right)}{{\hat{\eta }}_{s}}-\bm{\nabla} \cdot \bm{P}'\left( {{{\hat{\eta }}}_{s}},{{\Pi }_{s}} \right)=0  \quad \text{in}\quad{{\Omega }_{s}},
\label{eq:Linear3}
\end{equation}
\begin{equation}
\bm{\nabla} \cdot {\bm{\Sigma }_{e}}\left( {{{\hat{\eta }}}_{e}} \right)=0    \quad \text{in}\quad{{\Omega }_{e}},
\label{eq:Linear4}
\end{equation}
\begin{equation}
{{\hat{\eta }}_{e}}-{{\hat{\eta }}_{s}}=0   \quad \text{on} \quad   \Gamma.
\label{eq:Linear5}
\end{equation}

The flow equations {\color{blue}(12)-(13)} involve the linearized deformation operator $\bm{\Phi }'\left( {{{\hat{\eta }}}_{e}} \right)$ expressed in the steady deformed configuration as
\begin{equation}
\bm{\Phi}' \left( {{{\hat{\eta }}}_{e}} \right)=\left( \bm{\nabla} \cdot {{{\hat{\eta }}}_{e}} \right)I-\bm{\nabla} {{\hat{\eta }}_{e}}.
\end{equation}

In Eq. {\color{blue}(13)}, $\bm{\sigma}$ is the classical Cauchy stress tensor as
\begin{equation}
\bm{\sigma} \left( \hat{u},\hat{p} \right)=-\hat{p}\bm{I}+\frac{1}{{Re}}\left[ \bm{\nabla} \hat{\bm{u}}+{{\left( \bm{\nabla} \hat{\bm{u}} \right)}^{T}} \right],
\label{eq:pertubation_variables}
\end{equation}
while the linearization of the fluid stress tensor $\bm{\Sigma}'$ with respect to the extension displacement field  in the steady deformed configuration can be written as
\begin{equation}
\bm{\Sigma}' \left( \hat{\eta}_e ,U,P \right)=\bm{\sigma} \left( U,P \right)\bm{\Phi }'{{\left( {{\eta }_{e}} \right)}^{T}}-\frac{1}{{Re}}\left[ \left( \bm{\nabla} U \right)\left( \bm{\nabla} {\hat{\eta }_{e}} \right)+{{\left( \bm{\nabla} {\hat{\eta }_{e}} \right)}^{T}}{{\left( \bm{\nabla} U \right)}^{T}} \right].
\label{eq:eigenvalue_problem1}
\end{equation}

In the solid equation {\color{blue}(14)}, $\bm{P}'\left( {{{\hat{\eta }}}_{s}},{{\Pi }_{s}} \right)$ denotes the linearized first Piola-Kirchhoff stress tensor that is written in the steady deformed configuration as
\begin{equation}
\bm{P}'\left( {{{\hat{\eta }}}_{s}},{{\Pi }_{s}} \right)=\frac{1}{J\left( {{\Pi }_{s}} \right)}\left[ \bm{\nabla} {{{\hat{\eta }}}_{s}}\hat{\bm{F}}\left( {{\Pi }_{s}} \right)\hat{\bm{S}}\left( {{\Pi }_{s}} \right)+\hat{\bm{F}}\left( {{\Pi }_{s}} \right)\hat{\bm{S}}\left( \hat{\eta}_s,{{\Pi }_{s}} \right) \right]\hat{\bm{F}}\left( {{\Pi }_{s}} \right)^T,
\end{equation}
where the linearization of the second Piola-Kirchhoff stress tensor $\bm{S}\left( {{\eta }_{s}},{{\Pi }_{s}} \right)$ are
\begin{equation}
\begin{aligned}
  & \bm{S}\left( {{\eta }_{s}},{{\Pi }_{s}} \right)={{\lambda }_{s}}tr\left[ \bm{E}\left( {\hat{\eta }_{s}},{{\Pi }_{s}} \right) \right]+2{{\mu }_{s}}\bm{E}\left( {\hat{\eta }_{s}},{{\Pi }_{s}} \right), \\ 
 & \bm{E}\left( {{\eta }_{s}},{{\Pi }_{s}} \right)=\frac{1}{2}\bm{F}\left( {{\Pi }_{s}} \right)\left[ \bm{\nabla} \hat{\eta}_s +{{\left( \bm{\nabla} \hat{\eta}_s  \right)}^{T}} \right]F\left( {{\Pi }_{s}} \right). \\ 
\end{aligned}
\label{eq:eigenvalue_problem5}
\end{equation}

Equations {\color{blue}(15)} and {\color{blue}(16)} are the linearization of the extension sub-problem. At last, boundary conditions for the linearized FSI system are
\begin{equation}
\hat{\bm{u}}-{{\hat{\bm{u}}}_{s}}=0\quad\text{    on }\quad\Gamma, 
\end{equation}
\begin{equation}
\left[ \bm{\sigma} \left( \hat{\bm{u}},\hat{p} \right)+\bm{\Sigma }'\left( {{{\hat{\eta }}}_{e}},U,P \right) \right]\bm{n}-\bm{P}'\left( {\hat{\eta }_{s}},{{\Pi }_{s}} \right)\bm{n}=0\quad\text{    on }\quad\Gamma.
\end{equation}

Eq. {\color{blue}(12)-(16)} together with the boundary conditions {\color{blue}(22)-(23)} can be written in a form of generalized eigenvalue problem (EVP) \cite{jiang2022global, jiang2022instability}. To numerically solve this EVP, we use a shift-and-invert Arnoldi method from the ARPACK library \cite{lehoucq1998arpack}.
\subsection{Proper orthogonal decomposition method}
\label{subsec:RL}

We use the proper orthogonal decomposition to analyze the transition growth of nonlinear perturbation velocity $\tilde{\bm{{u}}}\left( \bm{x},t \right)=\bm{u}\left( \bm{x},t \right)-U\left( x \right),$ where velocity field is decomposed into a base flow $U$ and the nonlinear perturbation $\tilde{\bm{u}}$. POD analysis aims to find the optimal basis vectors ${{\phi }_{j}}\left( \bm{x} \right)$ that can best represent the given data $\tilde{\bm{u}}\left( \bm{x},t \right)$:
\begin{equation}
\tilde{\bm{u}}\left( \bm{x},t \right)=\sum\limits_{j}{{{a}_{j}}\left( t \right){{\phi }_{j}}\left( \bm{x} \right)},
\end{equation}
where ${{a}_{j}}\left( t \right)$ is temporal coefficients. The solution to this problem can be determined by finding the eigenvectors $\phi_j$ and the eigenvalues ${{\sigma }_{j}}$ from \cite{eckart1936approximation}
\begin{equation}
R{{\phi }_{j}}={{\sigma }_{j}}{{\phi }_{j}},\text{  }{{\phi }_{j}}\in {{\mathbb{R}}^{n}},\text{ }{{\sigma }_{1}}\ge \cdots \ge {{\sigma }_{n}}\ge 0,
\end{equation}
where $R$ is the covariance matrix of vector ${x}(t)$
\begin{equation}
	R=\sum\limits_{i=1}^{m}{x\left( {{t}_{i}} \right){{x}^{T}}\left( {{t}_{i}} \right)}=X{{X}^{T}}\in {{\mathbb{R}}^{n\times n}},
\end{equation}
where the matrix $X$ represents the $n$ snapshot data being stacked into a matrix form of
\begin{equation}
X=\left[ {\tilde{u}_{0}},{\tilde{u}_{1}},\cdots ,{\tilde{u}_{m-1}} \right]\in {{\mathbb{R}}^{n\times m}},
\end{equation}
where $m$ is the total number of grid points multiplied by the number of velocity components. The eigenvectors found in Eq. {\color{blue}(25)} are POD modes, which are orthonormal and satisfies
\begin{equation}
\left\langle {{\phi }_{j}},{{\phi }_{k}} \right\rangle =\int{{{\phi }_{j}}\cdot {{\phi }_{j}}d\Omega ={{\delta }_{jk}}}, \quad j,k=1,\cdots ,n.
\end{equation}

The eigenvalues corresponding to the POD modes are ranked from largest to smallest in order of importance in capturing the kinetic energy of the flow field. We can use the eigenvalues to determine the number of modes needed to represent the fluctuations in the flow field data. Generally, we retain only $r$ modes to express the flow such that
\begin{equation}
{\sum\limits_{j=1}^{r}{{{\sigma }_{j}}}}/{\sum\limits_{j=1}^{n}{{{\sigma }_{j}}}}\;\approx 1.
\end{equation}

We effectively reduce the high-dimensional $(n)$ flow field to be represented only with $r$ modes. Moreover, the temporal coefficients are determined accordingly by
\begin{equation}
{{a}_{j}}\left( t \right)=\left\langle \tilde{u}{{\left( x,t \right)}_{j}},{{\phi }_{j}}\left( x \right) \right\rangle =\left\langle \tilde{u}{{\left( t \right)}_{j}},{{\phi }_{j}} \right\rangle.
\end{equation}

\section{RESULTS AND DISCUSSION}
\label{sec:results}

\subsection{Bifurcation diagrams and hydrodynamics}
\label{subsec:GIA_FOM}

\begin{figure}[H]
\centerline{\includegraphics[width=0.85\linewidth]{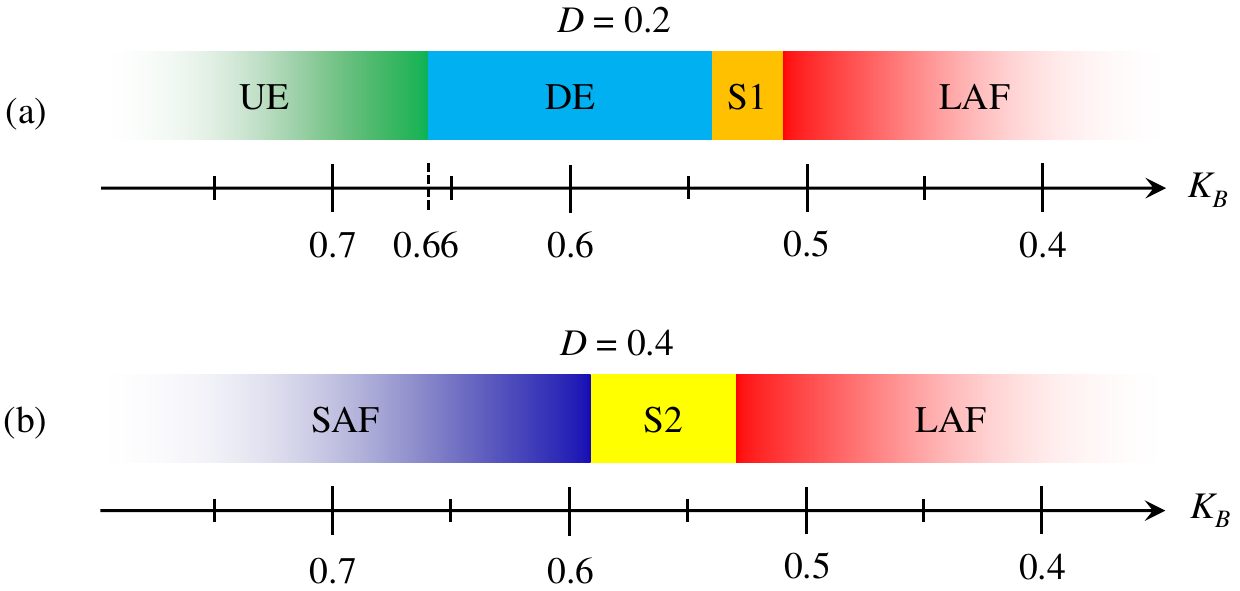}}
\caption{Bifurcation diagrams of the inverted-flag dynamics obtained by decreasing the stiffness $(K_B)$ at $Re = 200$. (a) $D = 0.2$ with UE: undeformed equilibrium; DE: deformed equilibrium; S1: small-deflection flapping (type1); and LAF: large-amplitude flapping. (b) $D = 0.4$ with SAF: small-amplitude flapping; S2: small-deflection flapping (type2).}
\label{fig:Fig2}
\end{figure}

This section contains an overview of the various regimes that are studied. We first use phase diagrams to illustrate the equilibria (and their stability) along with the flapping dynamics. Figure \ref{fig:Fig2} shows phase diagrams of flow patterns after direct numerical simulation as the decreasing stiffness $K_B$ from 0.9 to 0.3 for diameters $D = 0.2$ and 0.4 at $Re = 200$. Over 100 numerical simulations are carried out with an interval $\Delta K_B=-0.01$. For $D = 0.2$, the system transitions from the undeformed equilibrium (UE) to a stable deformed equilibrium (DE) when $K_B$ decreases to less than 0.66. Following this, the system bifurcates to small-deflection flapping (S1) when $K_B < 0.54$, and finally to the large-amplitude flapping (LAF) when $K_B$ is less than 0.52.

However, there is a different transition route for the larger cylinder case, i.e., $D = 0.4$. Phase of undeformed equilibrium and deformed equilibrium vanish, replaced by the small-amplitude flapping (SAF). Continuing to decrease $K_B$, a small-deflection flapping occurs when $K_B < 0.58$, where its behavior differs from that of $D = 0.2$ case and  we call it S2. We discuss the difference of hydrodynamic characteristic and instability mechanism of these two small-deflection flapping in Sec. {\color{blue}C}. At last, system enters the phase of large-amplitude flapping at $K_B < 0.56$. 
We find the existence of cylinder has weak effect on the performance of inverted flags system at the large-amplitude flapping phase. In this work, we focus on the onset instability of the elastic plate and hydrodynamics during these processes, and SAF is not considered.

\subsection{Linear instability analysis}
\label{subsec:BPOD_VIV}

\begin{figure}[H]
\centerline{\includegraphics[width=1\linewidth]{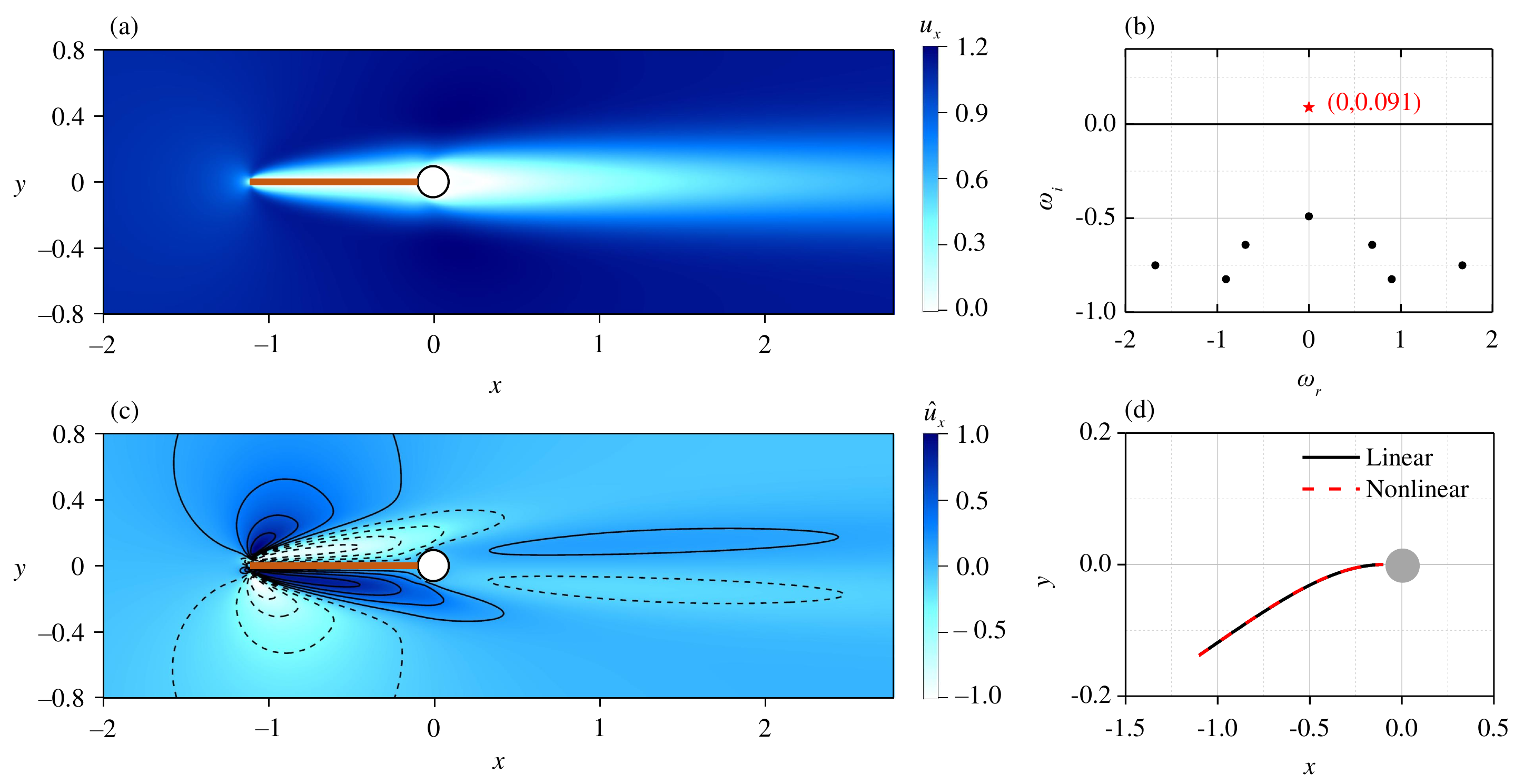}}
\caption{(a) Contour of the baseflow at $K_B = 0.6$ and $D = 0.2$. (b) Eigenvalue spectrum: red star indicates the unstable eigenmode; black circles represent stable eigenmodes. (c) shows streamwise velocity component of the leading eigenmode after normalization. (d) is the eigenvector distribution of plate displacement, where its magnitude is normed (red line) to compare the results of that calculated by DNS (black line).}
\label{fig:Fig3}
\end{figure}

We conduct the global instability analysis for case $D = 0.2$ and $K_B = 0.6$ to sight the instability mechanism. Figure \ref{fig:Fig3}(a) shows the spatial structure of the baseflow obtained by using the Newton method. The baseflow is converged to a symmetrical flow respect $y = 0$ and the streamwise deformation of the elastic plate is almost zero $O(10^{-3})$, i.e., the elastic plate is static undeformed equilibrium. Spectrum in Fig. \ref{fig:Fig3}(b) indicates there is a single steady ($\omega_r = 0$) unstable grow exponentially with a growing ratio $\omega_i = 0.091$ without spatial oscillation (pitchfork bifurcation). The governing operators are real valued, so the spectrum is necessarily symmetric with respect to the real axis \cite{golub2013matrix}. Figure \ref{fig:Fig3}(c) displays the contour for the leading eigenmode (streamwise velocity), where large value is found in the vicinity of the elastic plate and a weak long vortex located downstream of the cylinder. We call this type of eigenmode as “Structure mode” (SM) in this work. SM breaks the reflection symmetry of the steady flow, and for that reason it is called 'symmetry-breaking' shown in Fig. \ref{fig:Fig3}(d). The elastic plate is deflected downward here, but an upward deviation can be obtained by reversing the arbitrary sign of this mode. The bifurcation is mainly caused by the lifting of the elastic plate at the leading edge. Then, we compare the displacement of elastic plates at maximum deflection obtained from DNS with the result of linear analysis. There is a good match between these two curves after normalization, which indicates global linear instability analysis here can predict nonlinear behaviors of elastic plate. 

\begin{figure}[H]
\centerline{\includegraphics[width=1\linewidth]{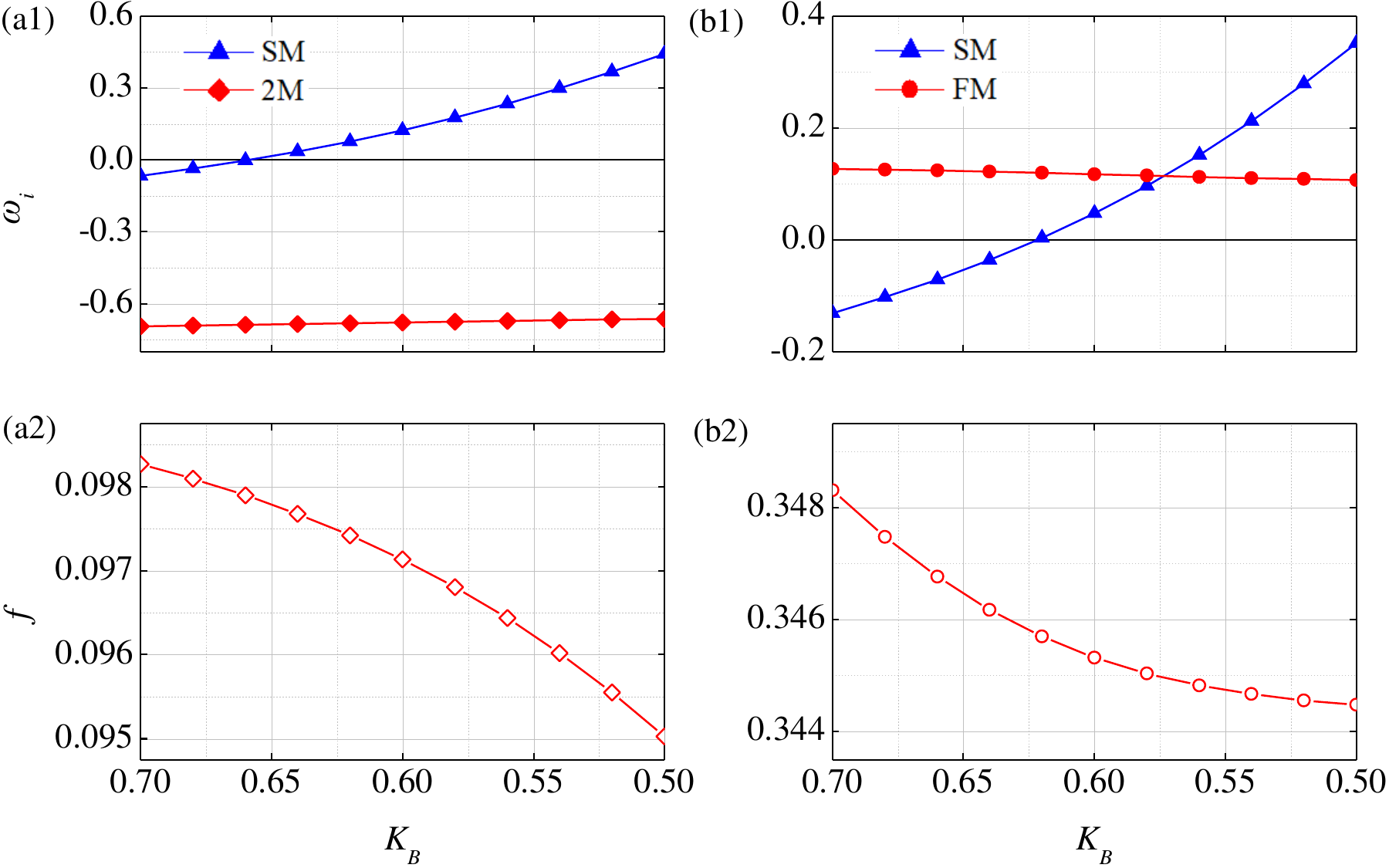}}
\caption{Eigenvalues vary with the decreasing of  $K_B$ from 0.7 to 0.5 at $Re = 200$ for cases: (a) $D = 0.2$ and (b) $D = 0.4$.}
\label{fig:Fig4}
\end{figure}

Besides the case of $K_B = 0.6$, we consider the dependence of eigenvalues on a variation of bending stiffness in a range of $K_B \in [0.5,0.7]$ at $D = 0.2$. SM is always the leading mode, which its growth ratio increases monotonously with the decrease of $K_B$ in Fig. \ref{fig:Fig4}(a1). When $K_B > 0.66$, the sign of $\omega_i$ changes from negative to positive corresponding to transition point shown in Fig. \ref{fig:Fig2}(a). The second mode 2M (ranking by their growth ratio) is a rapid damping mode, with its growth ratio near  $-0.62$. Moreover, different from Structure mode, 2M is an oscillating mode whose frequency decrease with the change of $K_B$ around 0.1, as shown in Fig. \ref{fig:Fig4}(a2). 2M contributes less for the first bifurcation (pitchfork bifurcation) but it is modified  with the deflection of elastic plate and contributes to the second bifurcation, i.e., the phase of deformed equilibrium transition to small-deflection flapping S1 \cite{goza2018global}. For larger cylinders, Fig. \ref{fig:Fig4}(b1) reveals the existence of two unstable modes: one is steady SM and the other is an oscillating mode ($\omega_r \neq 0$) at $D = 0.4$. We call this oscillating mode as “Fluid mode” (FM). Growth ratio of FM decreases with the decrease of $K_B$, which is almost independent on the variation of $K_B$. When $K_B > 0.57$, FM is the leading unstable mode and the type of bifurcation is judged as Hopf.

\begin{figure}[H]
\centerline{\includegraphics[width=1\linewidth]{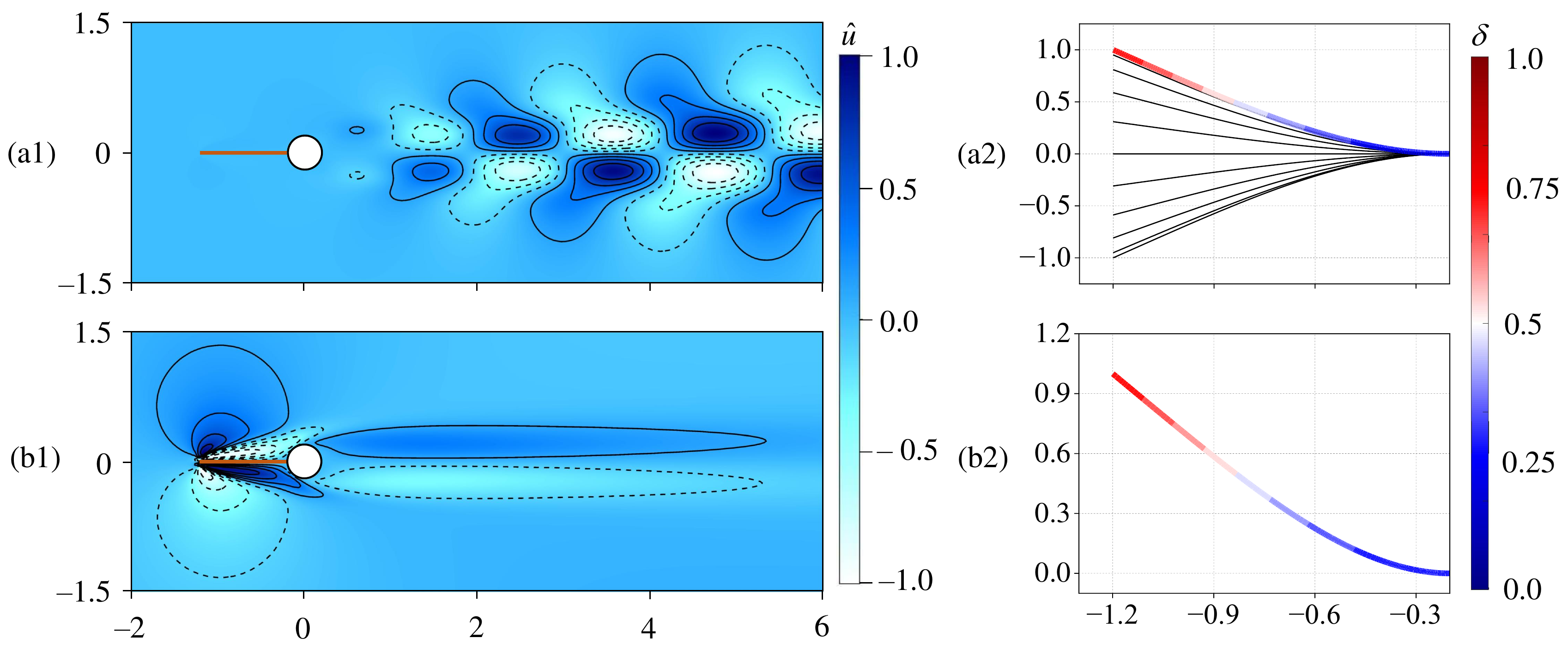}}
\caption{Contours of eigenvectors at $K_B = 0.6$ and $D = 0.4$ for (a) FM and (b) SM. (a1) and (b1) are streamwise velocities after normalization; (a2) and (b2) are displacement of the elastic plate normalized by the maximum tip displacement. Orange lines are eigenvectors, and grey line represent snapshots of the SAF phase obtained from DNS.}
\label{fig:Fig5}
\end{figure}

The eigenvector of FM demonstrates a typical vortex-shedding velocity pattern, i.e., alternating positive and negative streamwise velocities, which indicates the early stages of the unsteady Kármán vortex street development. This pattern is clearly visible in Fig. \ref{fig:Fig5}(a1). High values of $\hat{u}_x$ are primarily concentrated behind the cylinder, which distinguishes it from SM shown in Fig. \ref{fig:Fig5}(b1). Therefore, we conclude that FM and SM are decoupled, occupying different regions and having different instability mechanisms. The instability mechanism of FM is attributed to vortex shedding due to the presence of the cylinder, and bifurcation is a supercritical Hopf. Figs. \ref{fig:Fig5}(a2) and \ref{fig:Fig5}(b2) display the eigenvectors of the elastic plate (displacement after normalization) for FM and SM at $K_B = 0.6$, respectively. In this case, FM is the leading mode with $\omega_i = 0.14$, followed by SM. Although the unstable SM also contributes to the perturbation growth, the flow eventually develops into a symmetric, small-amplitude flapping, as depicted in Fig. \ref{fig:Fig5}(a2). Moreover, the linear mode's oscillation frequency, $f = 0.34$, is very close to the oscillation frequency of the nonlinear periodic solution, $f = 0.38$. The dynamical deformation is made clearer for the structure with a superposition of the plate’s position (the displacement being arbitrarily scaled) at different phases. We observe that vortex shedding downstream of the cylinder in FM can induce an oscillation of the elastic plate upward. Therefore, the instability mechanism in this case is attributed to vortex-induced vibration.

\begin{figure}[H]
\centerline{\includegraphics[width=0.76\linewidth]{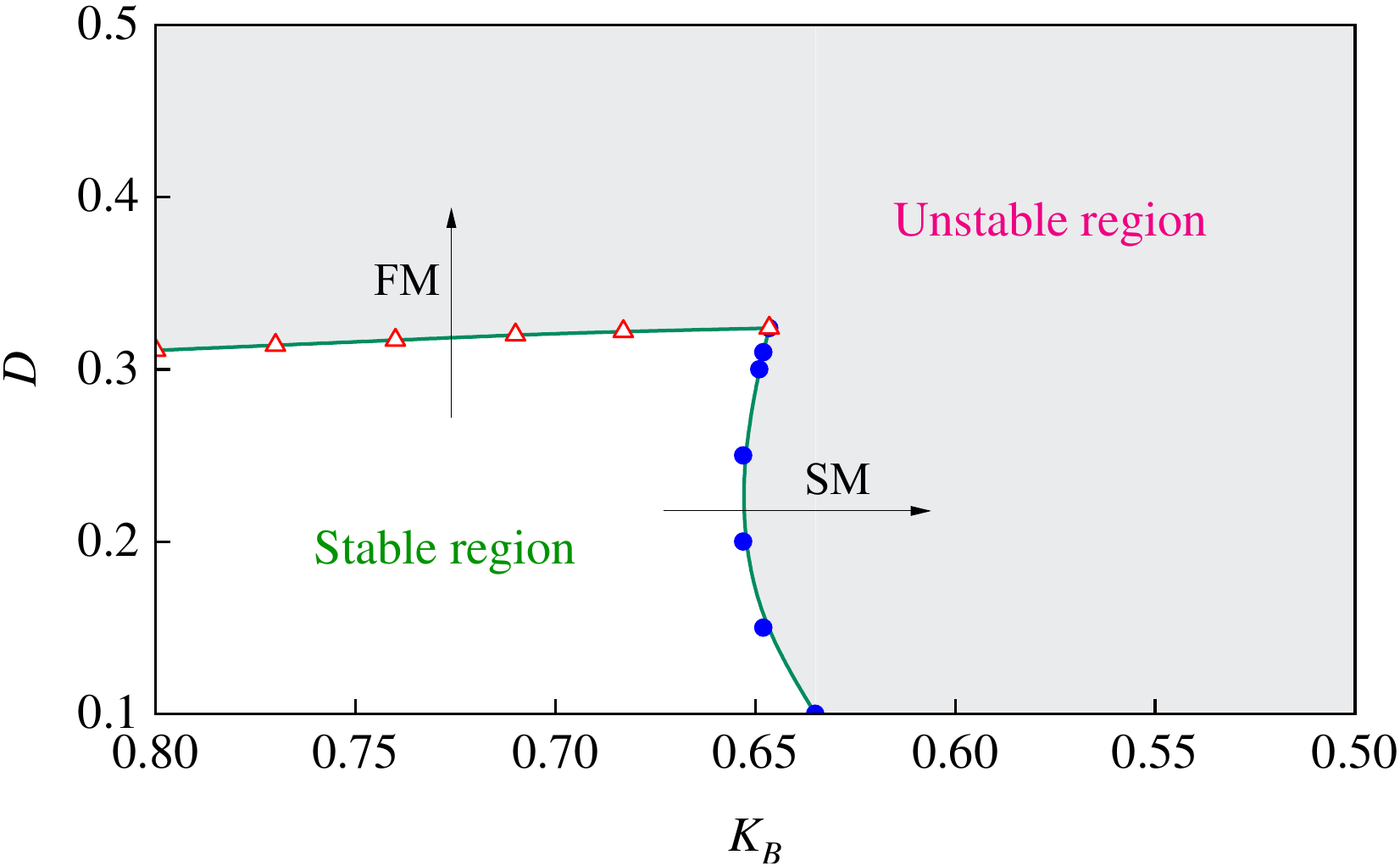}}
\caption{Neutral stability curve (green line) dividing flows as stable (white region) and unstable (grey region). Arrows are the growing direction ($\omega_i$) of the FM and SM branches.}
\label{fig:Fig6}
\end{figure}

After clarifying the information about the unstable eigenmodes, Fig. \ref{fig:Fig6} illustrates the neutral instability curve within the range of $(K_B,D) \in [0.8,0.5] \times[0.1,0.5]$. This curve is the intersection of the FM branch and the SM branch, and it divides the flow into stable and unstable regions. The stable region forms a rectangle in the lower left corner, while the remaining region is unstable. This means that the elastic plate remains in an undeformed equilibrium when a small cylinder or large bending stiffness is applied. The upper boundary of the rectangle is the FM branch, where the growth ratio is more sensitive to changes in diameter compared to bending stiffness. As a result, the FM branch is almost horizontal with respect to the $K_B$ axis. As the diameter is increased over the FM branch, the corresponding flow transitions from UE to SAF. The right boundary is the vertical SM branch, which is mainly influenced by the bending stiffness as shown in Fig. \ref{fig:Fig4}. UE transitions to the DE state with a decreasing $K_B$ while crossing the SM branch. The FM branch and SM branch intersect at the point $(K_B, D) = (0.647,0.324)$. We find that two-mode oscillating growth occurs near this point in the unstable region, which will be discussed in the next section.

\begin{figure}[H]
\centerline{\includegraphics[width=1\linewidth]{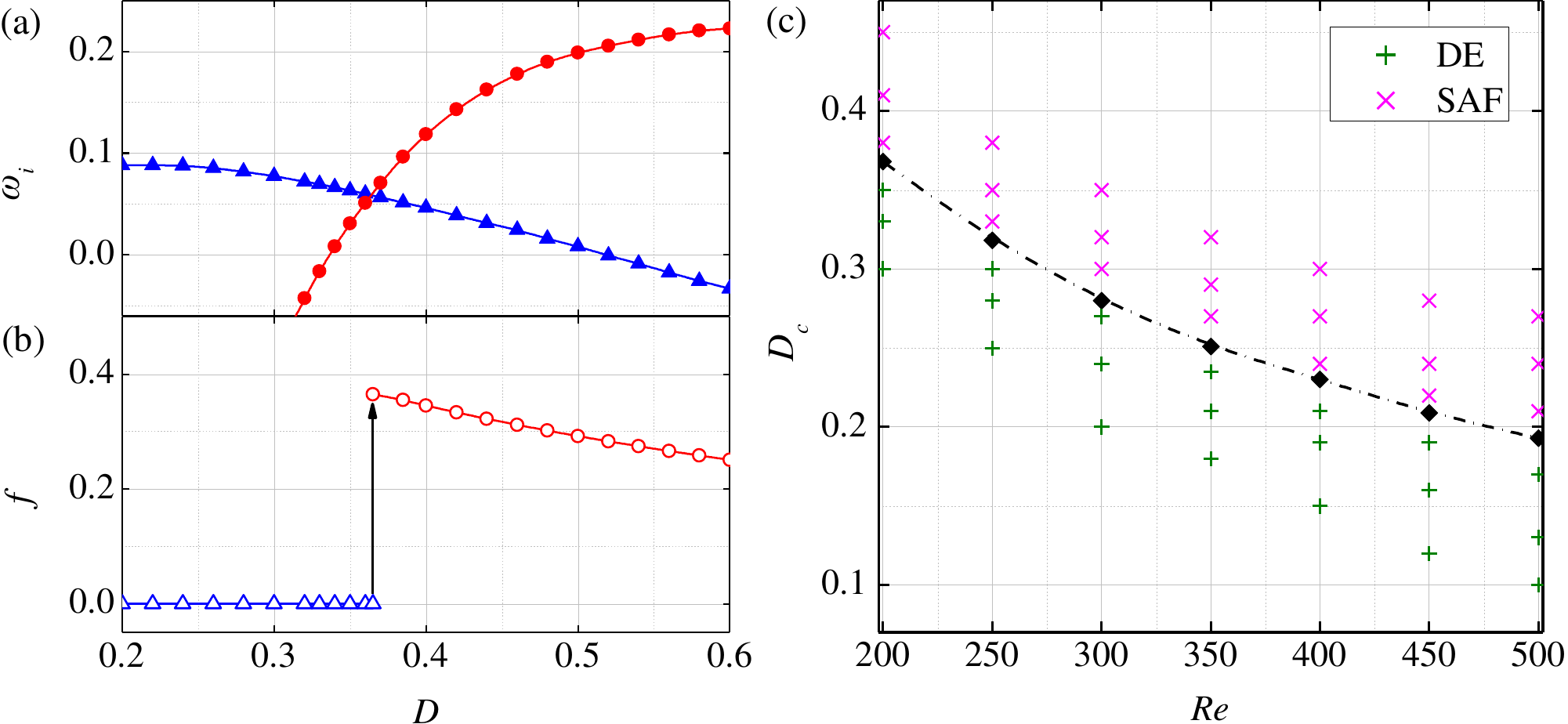}}
\caption{(a) and (b) Linear growth rate $\omega_i$ and frequency $f$ of the leading mode vary with the increase of dimeter in a range of $D \in [0.2,0.6]$ at $K_B = 0.6$. Flow transition from the  pitchfork bifurcation to the Hopf bifurcation when $D > 0.37$. (c) Critical transition diameter $D_c$ change with the increase of Reynold number from $Re = 200$ to 500. Times signs and plus signs represent the results of DNS.}
\label{fig:Fig7}
\end{figure}

The diameter of the cylinder affects the instability mechanism and transition performance of the inverted flags system. There is a critical diameter, denoted as $D_c$, which distinguishes different types of bifurcation from UE to DE/SAF. We aim to investigate the existence of $D_c$ and how the value of $D_c$ is influenced by the Reynolds number. In Figure \ref{fig:Fig7}(a), the variation of eigenvalues of the diameter from 0.2 to 0.6 at a fixed bending stiffness of $K_B = 0.6$ is depicted. The SM and FM represent the two leading unstable modes, exhibiting different sensitivities to changes in diameters. As the diameter ($D$) increases, the growth ratio $\omega_i$ of SM decreases monotonically from 0.09 at $D = 0.2$ to -0.04 at $D = 0.6$, while the trend of FM is the opposite. The critical diameter, $D_c$, is defined as the intersection (abscissa) of the two curves, and at this instance, $D_c = 0.37$. In Figure \ref{fig:Fig7}(b), the frequency of the most unstable eigenmodes is depicted. When $D < D_c$, the frequency is zero, and the bifurcation is a pitchfork. The bifurcation changes to Hopf when $D > D_c$. Next, we study the influence of the Reynolds number on the critical diameter in the range of $Re \in [200,500]$. Figure \ref{fig:Fig7}(c) illustrates that the critical diameter, $D_c$, decreases with the increase of $Re$ from 0.37 to 0.19. We anticipate that $D_c$ will continue to decrease with a growing Reynolds number. This result may be one reason why the UE and DE phases vanish when an elastic plate is clamped by a rather small cylinder in a high Reynolds number flow, as observed in the experiment conducted by Kim \emph{et al}. \cite{kim2013flapping}.

 \subsection{Transition growth and energy transfer}
\label{subsec:VIV_control}

After discussing the global linear instability, we focus on the initial transition growth of these unstable eigenmodes and the subsequent saturated oscillation during the nonlinear evolution. To investigate the nature of the bifurcation from UE to DE for small cylinder at $Re = 200$, we use the Landau equation, which expresses the evolution of the flow perturbation in its weakly nonlinear regime and it can be determined as \cite{guckenheimer2013nonlinear,aranson2002world}
\begin{equation}
\frac{dA}{dt}=\sigma \left( 1+i{{c}_{1}} \right)A-l\left( 1+i{{c}_{2}} \right){{\left| A \right|}^{2}}A+\cdot \cdot \cdot ,
\label{eq:Laudan}
\end{equation}
where parameter $\sigma$ indicates the linear growth rate of the perturbation and $l$ is the Landau constant. The sign of the $l$ determines the nature of the bifurcation, i.e., supercritical versus subcritical. The bifurcation is supercritical (non-hysteretic) if $l$ is positive ($l > 0$), while a negative value of $l$ ($l < 0$) indicates a subcritical (hysteretic) transition \cite{kang2017radial, kang2019numerical}. The sign of $l$ can be identified practically from the behavior of the instantaneous growth rate $dln|A|/dt$ as a function of $|A|^2$ at $|A|^2 \rightarrow 0$. The constants $c_1$ and $c_2$ are the linear and nonlinear dispersion coefficients, respectively. In this study, the coefficients are equal to zero ($c_1 = c_2 = 0$), because the critical modes are stationary $(\omega_r = 0)$ from linear instability analysis. The norm of the leading-edge transverse displacement $\delta$ is employed to define the amplitude of the perturbation $|A|$. 

\begin{figure}[H]
\centerline{\includegraphics[width=1\linewidth]{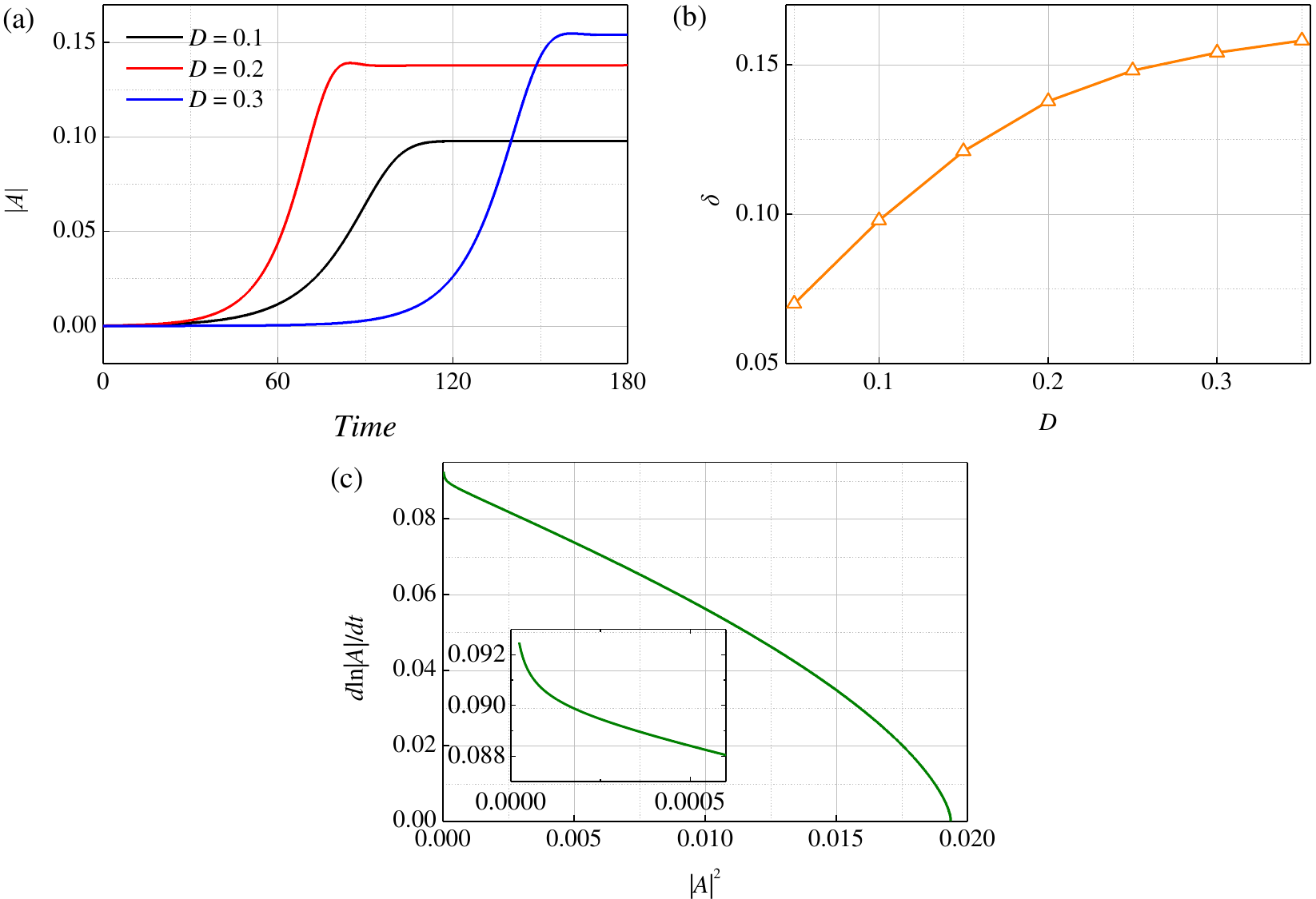}}
\caption{(a) The temporal evolution of the amplitude of the perturbation at $Re = 200$ with $K_B = 0.6$ for $D = 0.1, 0.2$, and 0.3. (b) The derivate of the amplitude logarithm plotted $d\text{ln}|A|/dt$ against the square of the amplitude $|A|^2$ for $D = 0.2$. The inset is the enlarge plot near $|A|^2 = 0$.}
\label{fig:Fig8}
\end{figure}

A time history of the amplitude $|A|$ at $Re = 200$ and $D = 0.1, 0.2$, and 0.3 is shown in Fig. \ref{fig:Fig8}(a). It can be observed that the amplitude of the perturbation $|A|$ exhibits exponential growth for these three cases, resulting in a growth rate $(\omega_i)$ for the most unstable mode, defined from the slope of the linear portion of the curve. As time progresses further, nonlinearity occurs at $t = 80$, and the amplitude slowly saturates to a constant value. Figure \ref{fig:Fig8}(b) displays the relationship between maximum transverse displacement $\delta$ and the diameter, indicating that $\delta$ increases with the diameter enlargement. The plot of $d\text{ln}|A|/dt$ versus $|A|^2$ at $D = 0.2$ is illustrated in Fig. \ref{fig:Fig8}(c). The intersection with the vertical axis provides the linear growth rate of the amplitude $|A|$, and the slope at the origin ($|A|^2 = 0$) determines the nonlinear bifurcation characteristics. The slope is negative, indicating that the type of bifurcation from UE to DE is supercritical for all examined diameters in phase DE according to Eq. \ref{eq:Laudan}. This conclusion is consistent with the result of Tavallaeinejad \emph{et al}. \cite{tavallaeinejad2018nonlinear} who studied inverted flags without considering the effect of the clamped cylinder. In summary, small cylinders do not change the type of bifurcation.

\begin{figure}[H]
\centerline{\includegraphics[width=1\linewidth]{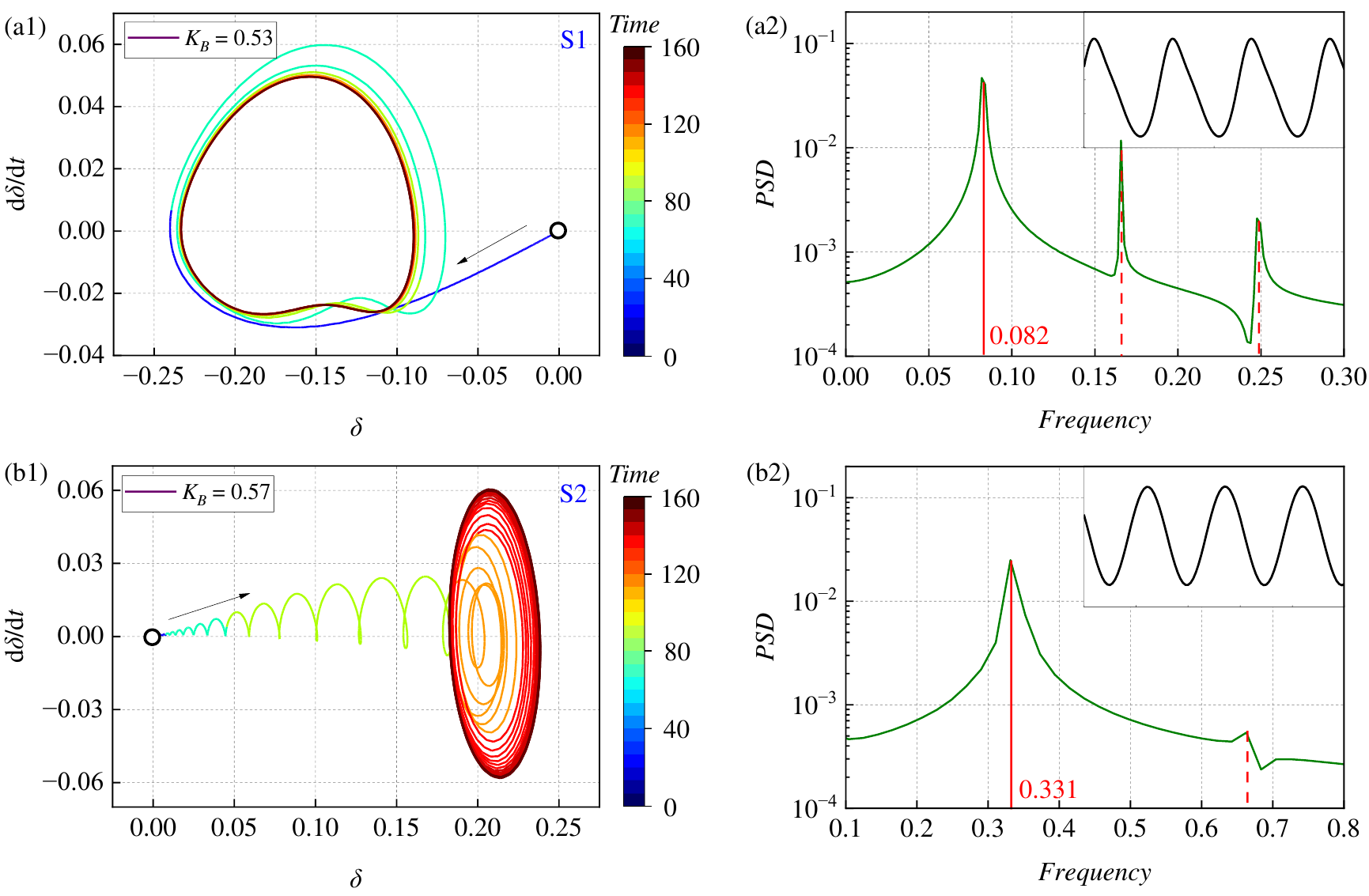}}
\caption{Dynamics characteristic of elastic plates for S1 (first row, $D = 0.2$ and $K_B = 0.53$) and S2 (second row, $D = 0.4$ and $K_B = 0.57$). (a1) and (b1) are phase trajectories of the maximum transverse tip displacement $\delta$. (a2) and (b2)are the FFT spectra of $\delta$ corresponding to these two cases. Insets are the temporal evolution of $\delta$. Fundamental frequencies are marked with the solid vertical line, noticeable harmonics with the dashed lines.}
\label{fig:Fig9}
\end{figure}

Then, we examined the hydrodynamic behavior of small-deflection flapping phases, S1 at $D = 0.2$ and S2 at $D = 0.4$. The simulations were initiated by introducing a steady undeformed equilibrium state. For S1, in Figure \ref{fig:Fig9}(a1), the phase trajectories of the maximum transverse displacement $\delta$ for $K_B = 0.53$ are shown. Initially, the steady undeformed equilibrium is disrupted, causing the elastic plate to transversely move downward when $t < 40$, and then the flow transitions directly to the small-deflection flapping phase (attractor morphology) instead of reaching a steady deformed equilibrium phase. During the saturated oscillation phase ($t > 100$), the spectra of $\delta$ were calculated using fast Fourier transform (FFT) to analyze the frequency characteristics. The analysis revealed only a fundamental frequency $f = 0.082$ and its harmonics, indicating that the system reached a single-period oscillation. For S2, in Figure \ref{fig:Fig9}(b1), a more complex phase trajectory of $\delta$ at $K_B = 0.57$ is displayed. Multiple oscillation modes, characterized by the displacement of the elastic plate increasing with oscillation, were observed during the initial transition phase when $t < 80$. Subsequently, the flow transitioned to a sine-like single-periodic oscillation with a higher frequency $f = 0.331$ compared to S1, as shown in Figure \ref{fig:Fig9}(b2). However, the amplitude of the transverse tip displacement $\Delta {\delta} = 0.032$ was lower than in the case of S1.

\begin{figure}[H]
\centerline{\includegraphics[width=1\linewidth]{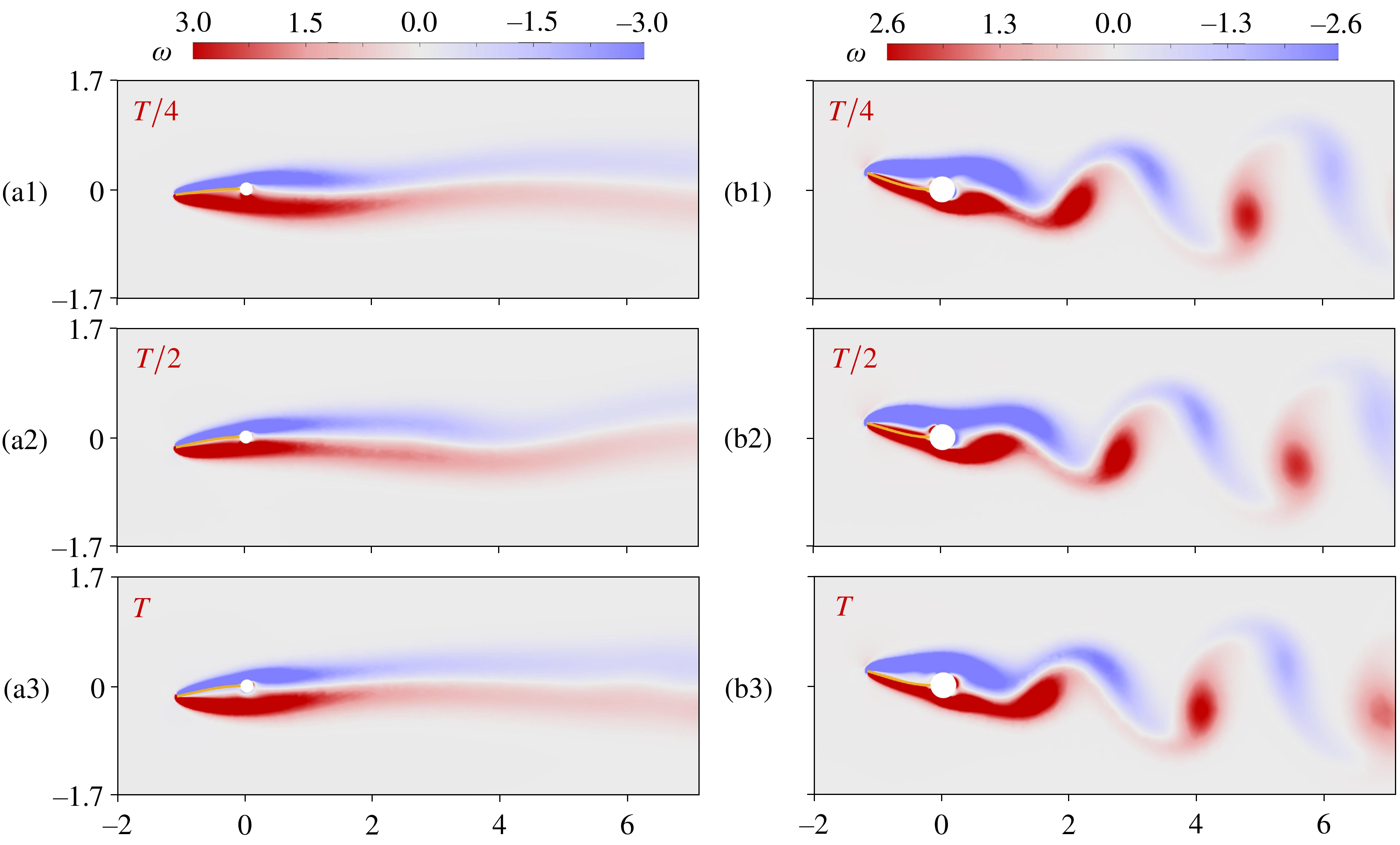}}
\caption{Snapshots of vorticity during a period for (a) S1 and (b) S2 phases.}
\label{fig:Fig10}
\end{figure}

To further explain the different instability mechanisms between these two types of small-deflection flapping (S1 and S2), let's take a look at three snapshots of vorticity during a period. In S1, as shown in Fig. \ref{fig:Fig10}(a), the leading vortex stays attached to the elastic plate and does not shed throughout the entire oscillating period, indicating that the dynamic performance of the elastic plates is not caused by classical vortex-induced vibrations (VIV)  \cite{goza2018global}. Gurugubelli and Jaiman also demonstrated that the leading-edge vortex formed during small-deflection flapping remains attached throughout the flapping process \cite{gurugubelli2015self}. On the other hand, vortex shedding occurs in S2, associated with the large cylinder, as shown in Fig. \ref{fig:Fig10}(b). We observe the occurrence of the von Kármán vortex street for S2, indicating VIV behavior. The point of vortex shedding is downstream of the cylinder rather than the elastic plate, and the leading-edge vortex remains attached to the plate throughout the flapping process. This leads us to a different conclusion from the instability mechanism LAF, suggesting that periodic flapping motion is caused by a complex interplay between the unsteady shedding of leading-edge vortices and the structural dynamics of the flexible flag, while the trailing-edge vortices have an insignificant effect on the flapping.

\begin{figure}[H]
\centerline{\includegraphics[width=1\linewidth]{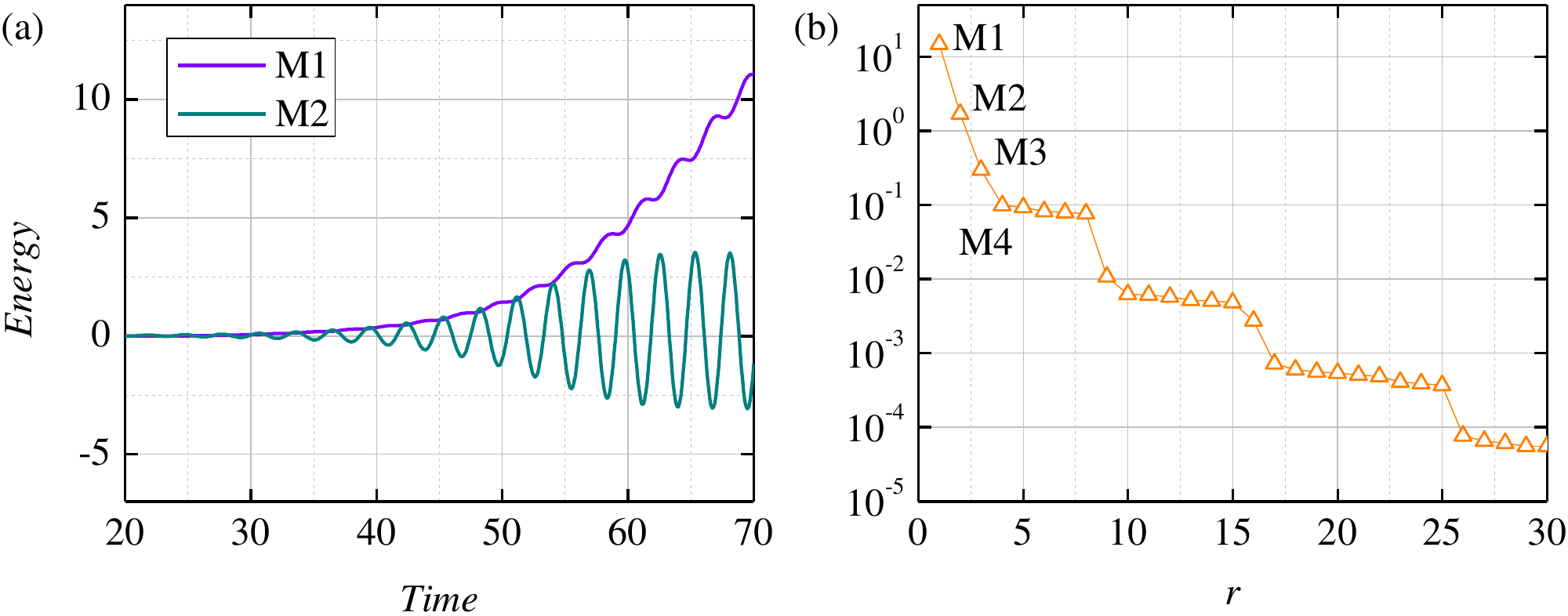}}
\caption{(a) Time-evolution of the first two POD modes. M1 and M2 represent the first and the second POD mode, respectively at $K_B = 0.57$ and $D = 0.4$. (b) The first 30 main singular values.}
\label{fig:Fig11}
\end{figure}

\begin{figure}[H]
\centerline{\includegraphics[width=1\linewidth]{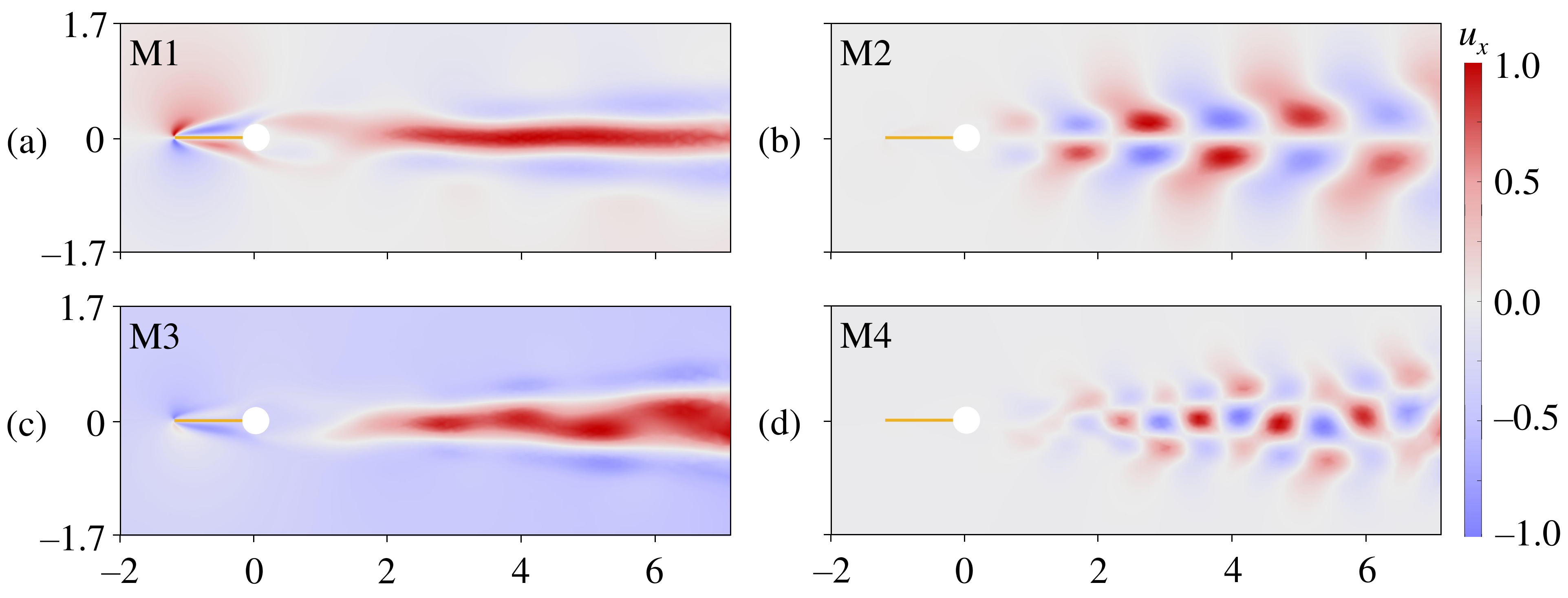}}
\caption{Contours of streamwise velocity of the first four POD modes.}
\label{fig:Fig12}
\end{figure}

Next, we analyze the multi-mode oscillation during the transition phase as shown in Fig. \ref{fig:Fig9}(b1) using proper orthogonal decomposition. A total of 500 velocity field snapshots, with a temporal spacing of 0.1, have been processed. These snapshots were taken from $Time = 20$ to 70 during the transition process. The number of snapshots $N$ is large enough to ensure well-converged POD results. In the ALE framework, the snapshots are collected at different grid positions, and the POD modes are plotted for a fixed grid. We have obtained a 50-order reduced order model, and the POD modes are ranked according to their energy. Figure \ref{fig:Fig11}(a) shows the time evolution of energy for the first two leading POD modes. The singular values of these modes are shown in Fig. \ref{fig:Fig11}(b), and we refer to the first four POD modes as M1 to M4. When $Time < 50$, M1 and M2 have the same amplitude and growth ratio, except M2 increases symmetrically with respect to the zero-energy line. After that, M2 tends to a saturated oscillation with a peak amplitude of 3.2. However, M1 continues to grow, reaching an energy level beyond 10 at $Time = 70$.

We display contours of streamwise velocity to further investigate the characteristics of these POD modes. Figure \ref{fig:Fig12}(a) and \ref{fig:Fig12}(b) show the spatial structure of the first two POD modes. We find their distribution are similar to the linear eigenmodes in Fig. \ref{fig:Fig5}(a1) and \ref{fig:Fig5}(b1). In this case, growth ratio of unstable FM is almost equal to SM, that is 0.115 $ vs$ 0.119, so these two decoupled modes can be equally excited. The first and second POD modes are believed to represent the domain coherent structures related to the lift and oscillation of the elastic plate, whereas the higher POD modes (Fig. \ref{fig:Fig12}(c) and \ref{fig:Fig12}(d)) reveal more intricate coherent structures, including secondary vortices and interacting vortices observed over and within the wake of the plate.

\begin{figure}[H]
\centerline{\includegraphics[width=0.75\linewidth]{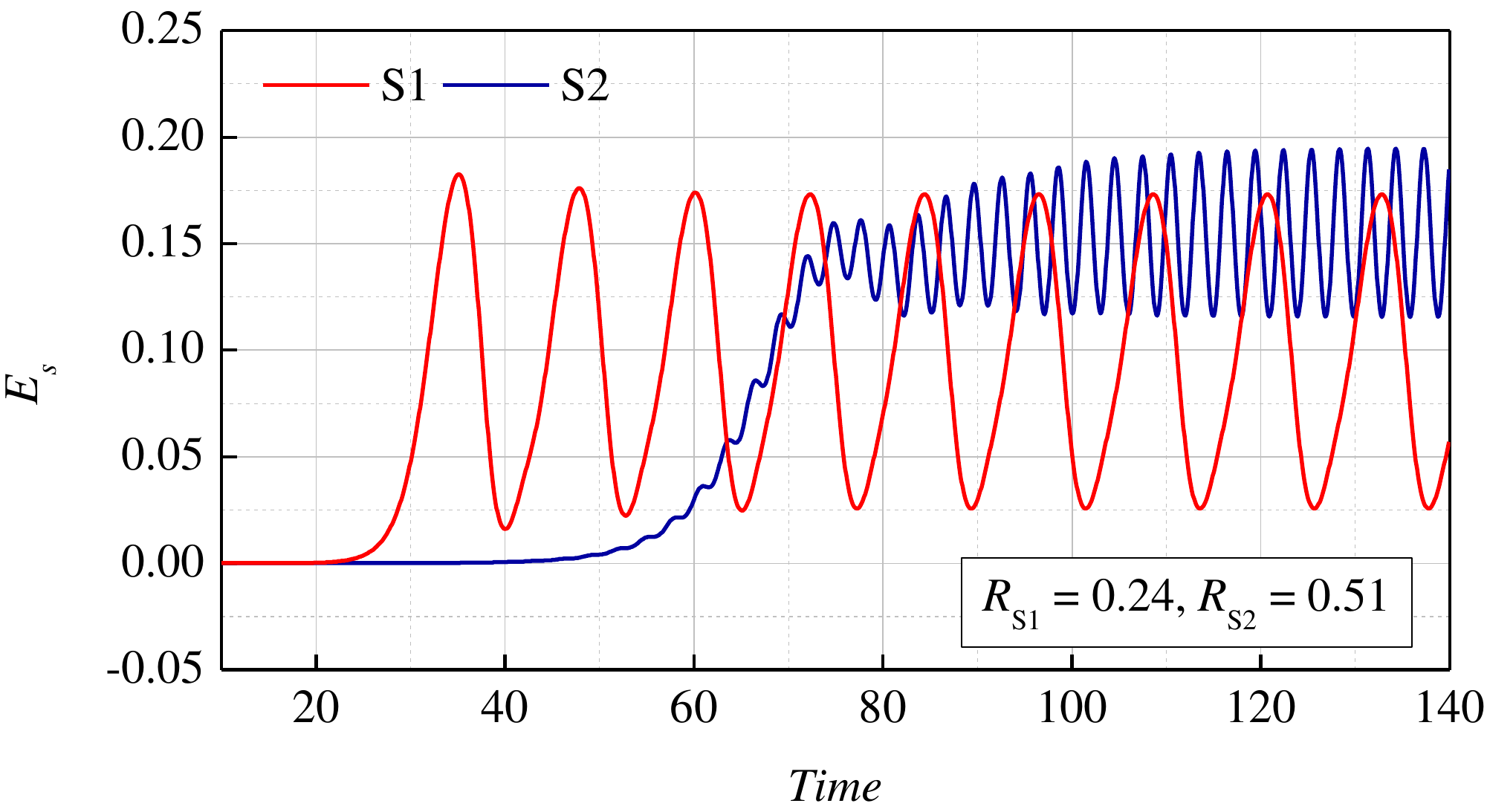}}
\caption{Time evolution of the bending strain energy $E_s$ for S1 ($K_B = 0.53$ and $D = 0.2$) and S2 ($K_B = 0.57$ and $D = 0.4$).}
\label{fig:Fig13}
\end{figure}

Finally, we analyze the ability of S1 and S2 extracting energy from the surrounding fluid flow. We define the non-dimensional bending strain energy $(E_s)$ and the ratio of the maximum strain energy to the total available fluid kinetic energy $(R)$ as \cite{gurugubelli2015self}, 
\begin{equation}
{{E}_{s}}=\frac{\int_{0}^{l}{EI{{k}^{2}}dl}}{2{{\rho }_{f}}U_{\infty }^{2}{{L}^{2}}},\text{ with }k=\frac{\left| {f}''\left( x \right) \right|}{{{\left[ 1+{f}'{{\left( x \right)}^{2}} \right]}^{{3}/{2}\;}}},
\end{equation}
\begin{equation}
R=\frac{{{\left( \Delta {{E}_{s}} \right)}_{max}}}{\frac{1}{2}{{\rho }_{f}}U_{\infty }^{2}\delta _{y}^{max}T},
\end{equation}
where $I$ is the moment of inertia; $k$ is the curvature of the deformed elastic plate; $f(x)$ is a piecewise polynomial function of sixth order that has been constructed to define the deformed plate profile; $f'$ and $f''$ are its first and second order derivatives, respectively. $(\Delta E_s)_{max}$ is the maximum change in the strain energy per oscillation; $\delta_y^{max}$ represents the maximum tip displacement and $T$ is the time taken to complete one half-cycle. Fig. \ref{fig:Fig13} gives the time evolution of non-dimensional bending strain energy $E_S$ form $Time = 0$ to 140 for S1 and S2, respectively. At the saturated oscillating phase, S2 has a higher effectiveness in extracting energy, that is, the mean value of $\bar{E}_s = 1.6$ and $\bar{E}_s = 1.0$ for S1 case. Moreover, due to the higher frequency oscillation of S2, the ratio of the maximum strain energy to the total available fluid kinetic energy for S2 is about twice for S1 case ($R_{S1} = 0.24$ and $R_{S2} = 0.51$). This result enlightens that we can obtain a efficient energy extracting by optimal the diameter of clamped cylinder.

\section{CONCLUSION}
\label{sec:conclusion}
In this work, we study numerically the onset instability of inverted flags clamped by a cylinder. The influence of bending stiffness $ K_B$, diameter of cylinder $D$, and Reynolds number $Re$ on transition are considered. We first give the bifurcation diagrams for diameter $D = 0.2$ and 0.4 by direct numerical simulations, and a total of six types of hydrodynamic phases are found at $Re = 200 $ and $K_B \in [0.9,0.4]$. Then, two unstable eigenmode branches: Structure mode SM and Flow mode FM are found by global linear instability analysis. FM is considered decoupled with SM, because FM concentrates on exciting perturbation at the wavemaker region downstream of the cylinder rather than the leading edge of the elastic plate of SM. The variation of SM branch is mainly dependent on the bending stiffness of the elastic plate and SM contributes to the pitchfork bifurcation (lifting instability mechanism) from the undeformed equilibrium (UE) to a stable deformed equilibrium (DE) for small cylinder cases. Moreover, the bifurcation is supercritical from the weakly nonlinear instability analysis. Larger cylinders cause the occurrence of unstable oscillating FM. When FM is the leading mode, flow transitions from UE to small amplitude flapping through a supercritical Hopf bifurcation. We also prove there is a critical diameter $D_c$ dividing the pitchfork bifurcation and Hopf bifurcation and $D_c$ decreases with the increase of Reynolds number monotonously.

Next, we use the proper orthogonal decomposition method to analyze the multi-modes transition growth due to the synergistic effect of unstable SM and FM at large cylinder cases. We calculate the energy transfer coefficients between two kinds of small deflected flapping corresponding to the small (S1) and large (S2) cylinder cases, respectively. A higher energy transfer effective from fluid to structure is found for S2. This work enlightens that we can prevent the instability of elastic plate or improve energy transfer efficiency of the inverted flags system by adjusting cylinder's diameter and bending stiffness in engineering.

\begin{acknowledgments}
The authors thanks for the valuable discussion with Dr. Jean-Lou Pfister on global stability analysis; gratefully acknowledge support from the Shenzhen Peacock Plan and the Cross-disciplinary Research and Innovation Fund provided by Tsinghua Shenzhen International Graduate School. 
\end{acknowledgments}

\vspace{0.5cm}
\textbf{Declaration of Interests}. The authors report no conflict of interest.

\setcounter{equation}{0}
\renewcommand\theequation{A\arabic{equation}} 

\section*{Appendix}

\label{sec:appendixA}
\subsection{Numerical method and verification of the nonlinear numerical solutions}

\begin{table}
\caption{{Maximum amplitude and flapping frequency of the inverted flag for $Re = 200$, and $K_B = 0.35$, as obtained using grid M1, M2, and M3.}}
\begin{ruledtabular}
\begin{tabular}{ccccc}
Mesh& M1 & M2 & M3& Goza \emph{et al.} \cite{goza2018global}.\\ \hline
Amplitude&0.79 &0.81 &0.81 &0.81  \\
Frequency&0.183 &0.179 &0.180 &0.180  \\
\end{tabular}
\end{ruledtabular}
\label{table:Tab1}
\end{table}

The spatial discretization of the governing equations {\color{blue}1-8} is obtained by a continuous Galerkin finite-element method. Quadratic (P2) elements are used to discretize to velocity and displacement fields while linear (P1) elements are used for the pressure and interface Lagrange multipliers. The corresponding finite-element bases are defined on unstructured meshes obtained after Delaunay triangulation of the computation domain. The extension region, is the same of the total fluid mesh. 

A fully implicit temporal scheme is then used to discretize in time the fluid-structure problem. More specifically, we use the Crank-Nicholson scheme proposed by Wick \cite{wick2012stability}, as it offers a good compromise between low dissipation and numerical stability. At each temporal iteration, the time-discretized fully nonlinear problem is solved by the Newton method with an exact Jacobin. The resulting sparse matrices and vectors are assembled using the open-source software FeniCS and the resolution of the linear system at each iteration of the Newton method is performed with the distributed direct sparse solver MUMPS \cite{alnaes2015fenics}. In this work, the time step size $\Delta t$ is determined by the Courant-Friedrichs-Lewy condition with Courant number being 0.1.

\begin{figure}[H]
\centerline{\includegraphics[width=1\linewidth]{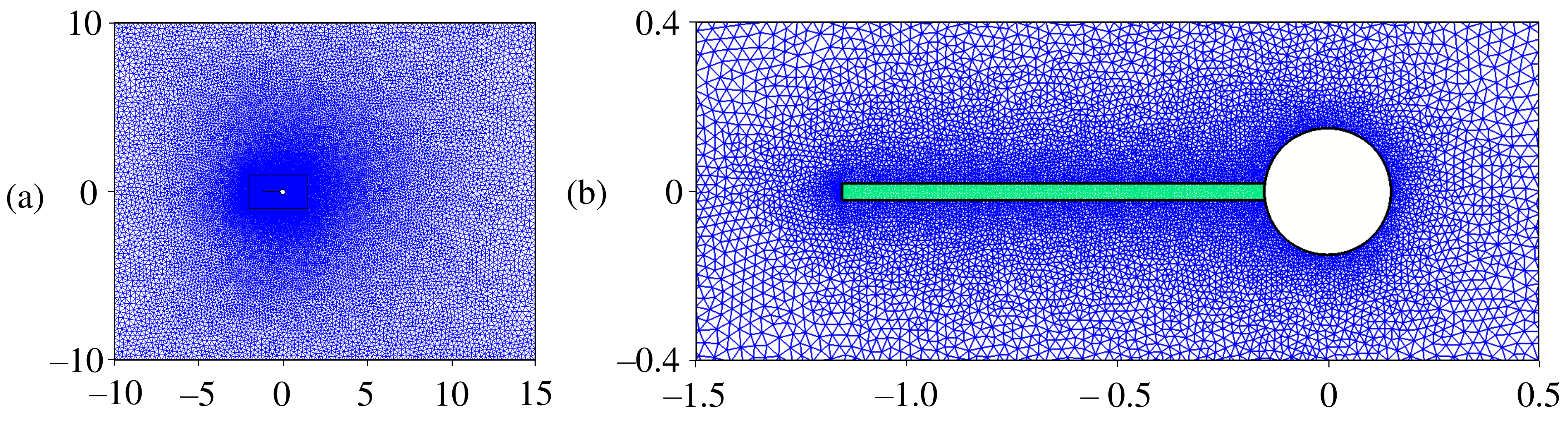}}
\caption{ (a) Finite-element mesh M2 for the complete fluid domain; (b) Close-up view of the central block and boundary layer mesh at the interface. The mesh in the solid region $\Omega_s$ is displayed in green while the discretization in the fluid region $\Omega_f$ is in blue. }
\label{fig:Fig14}
\end{figure}

We give a grid convergence study of the nonlinear solver for $Re = 200$, $M = 0.5$, and $K_B = 0.35$. For these parameters the flag enters limit-cycle flapping of fixed amplitude and frequency. The maximum amplitude and frequency of the elastic plate are shown in Table \ref{table:Tab1}. The results obtained with M2 and M3 have shown good agreement with the values given by Goza \emph{et al}., with an error less than $1\%$ \cite{goza2018global}. Figure \ref{fig:Fig14} shows the total computation mesh of M2, which is composed of 29976 triangles among which 1376 are located in the solid region. For the ease of computation, we use M2 mesh for DNS in this work.


\section*{References}
\bibliography{Drag_Reduction}

\begin{thebibliography}{44}%
\makeatletter
\providecommand \@ifxundefined [1]{%
 \@ifx{#1\undefined}
}%
\providecommand \@ifnum [1]{%
 \ifnum #1\expandafter \@firstoftwo
 \else \expandafter \@secondoftwo
 \fi
}%
\providecommand \@ifx [1]{%
 \ifx #1\expandafter \@firstoftwo
 \else \expandafter \@secondoftwo
 \fi
}%
\providecommand \natexlab [1]{#1}%
\providecommand \enquote  [1]{``#1''}%
\providecommand \bibnamefont  [1]{#1}%
\providecommand \bibfnamefont [1]{#1}%
\providecommand \citenamefont [1]{#1}%
\providecommand \href@noop [0]{\@secondoftwo}%
\providecommand \href [0]{\begingroup \@sanitize@url \@href}%
\providecommand \@href[1]{\@@startlink{#1}\@@href}%
\providecommand \@@href[1]{\endgroup#1\@@endlink}%
\providecommand \@sanitize@url [0]{\catcode `\\12\catcode `\$12\catcode
  `\&12\catcode `\#12\catcode `\^12\catcode `\_12\catcode `\%12\relax}%
\providecommand \@@startlink[1]{}%
\providecommand \@@endlink[0]{}%
\providecommand \url  [0]{\begingroup\@sanitize@url \@url }%
\providecommand \@url [1]{\endgroup\@href {#1}{\urlprefix }}%
\providecommand \urlprefix  [0]{URL }%
\providecommand \Eprint [0]{\href }%
\providecommand \doibase [0]{https://doi.org/}%
\providecommand \selectlanguage [0]{\@gobble}%
\providecommand \bibinfo  [0]{\@secondoftwo}%
\providecommand \bibfield  [0]{\@secondoftwo}%
\providecommand \translation [1]{[#1]}%
\providecommand \BibitemOpen [0]{}%
\providecommand \bibitemStop [0]{}%
\providecommand \bibitemNoStop [0]{.\EOS\space}%
\providecommand \EOS [0]{\spacefactor3000\relax}%
\providecommand \BibitemShut  [1]{\csname bibitem#1\endcsname}%
\let\auto@bib@innerbib\@empty
\bibitem [{\citenamefont {Zhang}\ \emph {et~al.}(2022)\citenamefont {Zhang},
  \citenamefont {Zhao}, \citenamefont {Huang}, \citenamefont {Lv},
  \citenamefont {Lu},\ and\ \citenamefont {Yu}}]{zhang2022effect}%
  \BibitemOpen
  \bibfield  {author} {\bibinfo {author} {\bibfnamefont {C.}~\bibnamefont
  {Zhang}}, \bibinfo {author} {\bibfnamefont {Z.}~\bibnamefont {Zhao}},
  \bibinfo {author} {\bibfnamefont {H.}~\bibnamefont {Huang}}, \bibinfo
  {author} {\bibfnamefont {X.}~\bibnamefont {Lv}}, \bibinfo {author}
  {\bibfnamefont {X.-Y.}\ \bibnamefont {Lu}},\ and\ \bibinfo {author}
  {\bibfnamefont {P.}~\bibnamefont {Yu}},\ }\bibfield  {title} {\bibinfo
  {title} {Effect of non-uniform stiffness distribution on the dynamics of
  inverted plates in a uniform flow},\ }\href@noop {} {\bibfield  {journal}
  {\bibinfo  {journal} {Physics of Fluids}\ }\textbf {\bibinfo {volume} {34}}
  (\bibinfo {year} {2022})}\BibitemShut {NoStop}%
\bibitem [{\citenamefont {Dong}\ \emph {et~al.}(2024)\citenamefont {Dong},
  \citenamefont {Tang}, \citenamefont {Zhao}, \citenamefont {Hu}, \citenamefont
  {Qu}, \citenamefont {Liu},\ and\ \citenamefont {Yang}}]{dong2024flag}%
  \BibitemOpen
  \bibfield  {author} {\bibinfo {author} {\bibfnamefont {L.}~\bibnamefont
  {Dong}}, \bibinfo {author} {\bibfnamefont {Q.}~\bibnamefont {Tang}}, \bibinfo
  {author} {\bibfnamefont {C.}~\bibnamefont {Zhao}}, \bibinfo {author}
  {\bibfnamefont {G.}~\bibnamefont {Hu}}, \bibinfo {author} {\bibfnamefont
  {S.}~\bibnamefont {Qu}}, \bibinfo {author} {\bibfnamefont {Z.}~\bibnamefont
  {Liu}},\ and\ \bibinfo {author} {\bibfnamefont {Y.}~\bibnamefont {Yang}},\
  }\bibfield  {title} {\bibinfo {title} {Flag-type hybrid nanogenerator
  utilizing flapping wakes for consistent high performance over an ultra-broad
  wind speed range},\ }\href@noop {} {\bibfield  {journal} {\bibinfo  {journal}
  {Nano Energy}\ }\textbf {\bibinfo {volume} {119}},\ \bibinfo {pages} {109057}
  (\bibinfo {year} {2024})}\BibitemShut {NoStop}%
\bibitem [{\citenamefont {Orrego}\ \emph {et~al.}(2017)\citenamefont {Orrego},
  \citenamefont {Shoele}, \citenamefont {Ruas}, \citenamefont {Doran},
  \citenamefont {Caggiano}, \citenamefont {Mittal},\ and\ \citenamefont
  {Kang}}]{orrego2017harvesting}%
  \BibitemOpen
  \bibfield  {author} {\bibinfo {author} {\bibfnamefont {S.}~\bibnamefont
  {Orrego}}, \bibinfo {author} {\bibfnamefont {K.}~\bibnamefont {Shoele}},
  \bibinfo {author} {\bibfnamefont {A.}~\bibnamefont {Ruas}}, \bibinfo {author}
  {\bibfnamefont {K.}~\bibnamefont {Doran}}, \bibinfo {author} {\bibfnamefont
  {B.}~\bibnamefont {Caggiano}}, \bibinfo {author} {\bibfnamefont
  {R.}~\bibnamefont {Mittal}},\ and\ \bibinfo {author} {\bibfnamefont {S.~H.}\
  \bibnamefont {Kang}},\ }\bibfield  {title} {\bibinfo {title} {Harvesting
  ambient wind energy with an inverted piezoelectric flag},\ }\href@noop {}
  {\bibfield  {journal} {\bibinfo  {journal} {Applied energy}\ }\textbf
  {\bibinfo {volume} {194}},\ \bibinfo {pages} {212} (\bibinfo {year}
  {2017})}\BibitemShut {NoStop}%
\bibitem [{\citenamefont {Silva-Leon}\ \emph {et~al.}(2019)\citenamefont
  {Silva-Leon}, \citenamefont {Cioncolini}, \citenamefont {Nabawy},
  \citenamefont {Revell},\ and\ \citenamefont
  {Kennaugh}}]{silva2019simultaneous}%
  \BibitemOpen
  \bibfield  {author} {\bibinfo {author} {\bibfnamefont {J.}~\bibnamefont
  {Silva-Leon}}, \bibinfo {author} {\bibfnamefont {A.}~\bibnamefont
  {Cioncolini}}, \bibinfo {author} {\bibfnamefont {M.~R.}\ \bibnamefont
  {Nabawy}}, \bibinfo {author} {\bibfnamefont {A.}~\bibnamefont {Revell}},\
  and\ \bibinfo {author} {\bibfnamefont {A.}~\bibnamefont {Kennaugh}},\
  }\bibfield  {title} {\bibinfo {title} {Simultaneous wind and solar energy
  harvesting with inverted flags},\ }\href@noop {} {\bibfield  {journal}
  {\bibinfo  {journal} {Applied Energy}\ }\textbf {\bibinfo {volume} {239}},\
  \bibinfo {pages} {846} (\bibinfo {year} {2019})}\BibitemShut {NoStop}%
\bibitem [{\citenamefont {Umair}\ \emph {et~al.}(2022)\citenamefont {Umair},
  \citenamefont {Latif}, \citenamefont {Uddin},\ and\ \citenamefont
  {Abdelkefi}}]{umair2022experimental}%
  \BibitemOpen
  \bibfield  {author} {\bibinfo {author} {\bibfnamefont {M.}~\bibnamefont
  {Umair}}, \bibinfo {author} {\bibfnamefont {U.}~\bibnamefont {Latif}},
  \bibinfo {author} {\bibfnamefont {E.}~\bibnamefont {Uddin}},\ and\ \bibinfo
  {author} {\bibfnamefont {A.}~\bibnamefont {Abdelkefi}},\ }\bibfield  {title}
  {\bibinfo {title} {Experimental hydrodynamic investigations on the
  effectiveness of inverted flag-based piezoelectric energy harvester in the
  wake of bluff body},\ }\href@noop {} {\bibfield  {journal} {\bibinfo
  {journal} {Ocean Engineering}\ }\textbf {\bibinfo {volume} {245}},\ \bibinfo
  {pages} {110454} (\bibinfo {year} {2022})}\BibitemShut {NoStop}%
\bibitem [{\citenamefont {Lim}\ and\ \citenamefont
  {Park}(2022)}]{lim2022numerical}%
  \BibitemOpen
  \bibfield  {author} {\bibinfo {author} {\bibfnamefont {S.~H.}\ \bibnamefont
  {Lim}}\ and\ \bibinfo {author} {\bibfnamefont {S.~G.}\ \bibnamefont {Park}},\
  }\bibfield  {title} {\bibinfo {title} {Numerical analysis of energy
  harvesting system including an inclined inverted flag},\ }\href@noop {}
  {\bibfield  {journal} {\bibinfo  {journal} {Physics of Fluids}\ }\textbf
  {\bibinfo {volume} {34}} (\bibinfo {year} {2022})}\BibitemShut {NoStop}%
\bibitem [{\citenamefont {Tavallaeinejad}\ \emph
  {et~al.}(2020{\natexlab{a}})\citenamefont {Tavallaeinejad}, \citenamefont
  {Pa{\"\i}doussis}, \citenamefont {Legrand},\ and\ \citenamefont
  {Kheiri}}]{tavallaeinejad2020instability}%
  \BibitemOpen
  \bibfield  {author} {\bibinfo {author} {\bibfnamefont {M.}~\bibnamefont
  {Tavallaeinejad}}, \bibinfo {author} {\bibfnamefont {M.~P.}\ \bibnamefont
  {Pa{\"\i}doussis}}, \bibinfo {author} {\bibfnamefont {M.}~\bibnamefont
  {Legrand}},\ and\ \bibinfo {author} {\bibfnamefont {M.}~\bibnamefont
  {Kheiri}},\ }\bibfield  {title} {\bibinfo {title} {Instability and the
  post-critical behaviour of two-dimensional inverted flags in axial flow},\
  }\href@noop {} {\bibfield  {journal} {\bibinfo  {journal} {Journal of Fluid
  Mechanics}\ }\textbf {\bibinfo {volume} {890}},\ \bibinfo {pages} {A14}
  (\bibinfo {year} {2020}{\natexlab{a}})}\BibitemShut {NoStop}%
\bibitem [{\citenamefont {Shen}\ \emph {et~al.}(2024)\citenamefont {Shen},
  \citenamefont {Zhao}, \citenamefont {Zhang}, \citenamefont {Wang},
  \citenamefont {Yang}, \citenamefont {Wei} \emph {et~al.}}]{shen2024coupled}%
  \BibitemOpen
  \bibfield  {author} {\bibinfo {author} {\bibfnamefont {P.}~\bibnamefont
  {Shen}}, \bibinfo {author} {\bibfnamefont {W.}~\bibnamefont {Zhao}}, \bibinfo
  {author} {\bibfnamefont {H.}~\bibnamefont {Zhang}}, \bibinfo {author}
  {\bibfnamefont {Z.}~\bibnamefont {Wang}}, \bibinfo {author} {\bibfnamefont
  {H.}~\bibnamefont {Yang}}, \bibinfo {author} {\bibfnamefont {Y.}~\bibnamefont
  {Wei}}, \emph {et~al.},\ }\bibfield  {title} {\bibinfo {title} {Coupled
  motion characteristic of standard flag and inverted flag in tandem
  arrangement},\ }\href@noop {} {\bibfield  {journal} {\bibinfo  {journal}
  {Experimental Thermal and Fluid Science}\ }\textbf {\bibinfo {volume}
  {150}},\ \bibinfo {pages} {111055} (\bibinfo {year} {2024})}\BibitemShut
  {NoStop}%
\bibitem [{\citenamefont {Cioncolini}\ \emph {et~al.}(2019)\citenamefont
  {Cioncolini}, \citenamefont {Nabawy}, \citenamefont {Silva-Leon},
  \citenamefont {O’connor},\ and\ \citenamefont
  {Revell}}]{cioncolini2019experimental}%
  \BibitemOpen
  \bibfield  {author} {\bibinfo {author} {\bibfnamefont {A.}~\bibnamefont
  {Cioncolini}}, \bibinfo {author} {\bibfnamefont {M.~R.}\ \bibnamefont
  {Nabawy}}, \bibinfo {author} {\bibfnamefont {J.}~\bibnamefont {Silva-Leon}},
  \bibinfo {author} {\bibfnamefont {J.}~\bibnamefont {O’connor}},\ and\
  \bibinfo {author} {\bibfnamefont {A.}~\bibnamefont {Revell}},\ }\bibfield
  {title} {\bibinfo {title} {An experimental and computational study on
  inverted flag dynamics for simultaneous wind--solar energy harvesting},\
  }\href@noop {} {\bibfield  {journal} {\bibinfo  {journal} {Fluids}\ }\textbf
  {\bibinfo {volume} {4}},\ \bibinfo {pages} {87} (\bibinfo {year}
  {2019})}\BibitemShut {NoStop}%
\bibitem [{\citenamefont {Hu}\ \emph {et~al.}(2020)\citenamefont {Hu},
  \citenamefont {Feng},\ and\ \citenamefont {Wang}}]{hu2020passive}%
  \BibitemOpen
  \bibfield  {author} {\bibinfo {author} {\bibfnamefont {Y.-W.}\ \bibnamefont
  {Hu}}, \bibinfo {author} {\bibfnamefont {L.-H.}\ \bibnamefont {Feng}},\ and\
  \bibinfo {author} {\bibfnamefont {J.-J.}\ \bibnamefont {Wang}},\ }\bibfield
  {title} {\bibinfo {title} {Passive oscillations of inverted flags in a
  uniform flow},\ }\href@noop {} {\bibfield  {journal} {\bibinfo  {journal}
  {Journal of Fluid Mechanics}\ }\textbf {\bibinfo {volume} {884}},\ \bibinfo
  {pages} {A32} (\bibinfo {year} {2020})}\BibitemShut {NoStop}%
\bibitem [{\citenamefont {Kim}\ \emph {et~al.}(2013)\citenamefont {Kim},
  \citenamefont {Coss{\'e}}, \citenamefont {Cerdeira},\ and\ \citenamefont
  {Gharib}}]{kim2013flapping}%
  \BibitemOpen
  \bibfield  {author} {\bibinfo {author} {\bibfnamefont {D.}~\bibnamefont
  {Kim}}, \bibinfo {author} {\bibfnamefont {J.}~\bibnamefont {Coss{\'e}}},
  \bibinfo {author} {\bibfnamefont {C.~H.}\ \bibnamefont {Cerdeira}},\ and\
  \bibinfo {author} {\bibfnamefont {M.}~\bibnamefont {Gharib}},\ }\bibfield
  {title} {\bibinfo {title} {Flapping dynamics of an inverted flag},\
  }\href@noop {} {\bibfield  {journal} {\bibinfo  {journal} {Journal of Fluid
  Mechanics}\ }\textbf {\bibinfo {volume} {736}},\ \bibinfo {pages} {R1}
  (\bibinfo {year} {2013})}\BibitemShut {NoStop}%
\bibitem [{\citenamefont {Sader}\ \emph
  {et~al.}(2016{\natexlab{a}})\citenamefont {Sader}, \citenamefont {Coss{\'e}},
  \citenamefont {Kim}, \citenamefont {Fan},\ and\ \citenamefont
  {Gharib}}]{sader2016large}%
  \BibitemOpen
  \bibfield  {author} {\bibinfo {author} {\bibfnamefont {J.~E.}\ \bibnamefont
  {Sader}}, \bibinfo {author} {\bibfnamefont {J.}~\bibnamefont {Coss{\'e}}},
  \bibinfo {author} {\bibfnamefont {D.}~\bibnamefont {Kim}}, \bibinfo {author}
  {\bibfnamefont {B.}~\bibnamefont {Fan}},\ and\ \bibinfo {author}
  {\bibfnamefont {M.}~\bibnamefont {Gharib}},\ }\bibfield  {title} {\bibinfo
  {title} {Large-amplitude flapping of an inverted flag in a uniform steady
  flow--a vortex-induced vibration},\ }\href@noop {} {\bibfield  {journal}
  {\bibinfo  {journal} {Journal of Fluid Mechanics}\ }\textbf {\bibinfo
  {volume} {793}},\ \bibinfo {pages} {524} (\bibinfo {year}
  {2016}{\natexlab{a}})}\BibitemShut {NoStop}%
\bibitem [{\citenamefont {Fan}\ \emph {et~al.}(2019)\citenamefont {Fan},
  \citenamefont {Huertas-Cerdeira}, \citenamefont {Coss{\'e}}, \citenamefont
  {Sader},\ and\ \citenamefont {Gharib}}]{fan2019effect}%
  \BibitemOpen
  \bibfield  {author} {\bibinfo {author} {\bibfnamefont {B.}~\bibnamefont
  {Fan}}, \bibinfo {author} {\bibfnamefont {C.}~\bibnamefont
  {Huertas-Cerdeira}}, \bibinfo {author} {\bibfnamefont {J.}~\bibnamefont
  {Coss{\'e}}}, \bibinfo {author} {\bibfnamefont {J.~E.}\ \bibnamefont
  {Sader}},\ and\ \bibinfo {author} {\bibfnamefont {M.}~\bibnamefont
  {Gharib}},\ }\bibfield  {title} {\bibinfo {title} {Effect of morphology on
  the large-amplitude flapping dynamics of an inverted flag in a uniform
  flow},\ }\href@noop {} {\bibfield  {journal} {\bibinfo  {journal} {Journal of
  Fluid Mechanics}\ }\textbf {\bibinfo {volume} {874}},\ \bibinfo {pages} {526}
  (\bibinfo {year} {2019})}\BibitemShut {NoStop}%
\bibitem [{\citenamefont {Yu}\ \emph {et~al.}(2017)\citenamefont {Yu},
  \citenamefont {Liu},\ and\ \citenamefont {Chen}}]{yu2017vortex}%
  \BibitemOpen
  \bibfield  {author} {\bibinfo {author} {\bibfnamefont {Y.}~\bibnamefont
  {Yu}}, \bibinfo {author} {\bibfnamefont {Y.}~\bibnamefont {Liu}},\ and\
  \bibinfo {author} {\bibfnamefont {Y.}~\bibnamefont {Chen}},\ }\bibfield
  {title} {\bibinfo {title} {Vortex dynamics behind a self-oscillating inverted
  flag placed in a channel flow: Time-resolved particle image velocimetry
  measurements},\ }\href@noop {} {\bibfield  {journal} {\bibinfo  {journal}
  {Physics of Fluids}\ }\textbf {\bibinfo {volume} {29}} (\bibinfo {year}
  {2017})}\BibitemShut {NoStop}%
\bibitem [{\citenamefont {Ryu}\ \emph {et~al.}(2015)\citenamefont {Ryu},
  \citenamefont {Park}, \citenamefont {Kim},\ and\ \citenamefont
  {Sung}}]{ryu2015flapping}%
  \BibitemOpen
  \bibfield  {author} {\bibinfo {author} {\bibfnamefont {J.}~\bibnamefont
  {Ryu}}, \bibinfo {author} {\bibfnamefont {S.~G.}\ \bibnamefont {Park}},
  \bibinfo {author} {\bibfnamefont {B.}~\bibnamefont {Kim}},\ and\ \bibinfo
  {author} {\bibfnamefont {H.~J.}\ \bibnamefont {Sung}},\ }\bibfield  {title}
  {\bibinfo {title} {Flapping dynamics of an inverted flag in a uniform flow},\
  }\href@noop {} {\bibfield  {journal} {\bibinfo  {journal} {Journal of Fluids
  and Structures}\ }\textbf {\bibinfo {volume} {57}},\ \bibinfo {pages} {159}
  (\bibinfo {year} {2015})}\BibitemShut {NoStop}%
\bibitem [{\citenamefont {Cao}\ \emph {et~al.}(2021)\citenamefont {Cao},
  \citenamefont {Wang},\ and\ \citenamefont {Wang}}]{cao2021spatially}%
  \BibitemOpen
  \bibfield  {author} {\bibinfo {author} {\bibfnamefont {S.}~\bibnamefont
  {Cao}}, \bibinfo {author} {\bibfnamefont {G.}~\bibnamefont {Wang}},\ and\
  \bibinfo {author} {\bibfnamefont {K.~G.}\ \bibnamefont {Wang}},\ }\bibfield
  {title} {\bibinfo {title} {A spatially varying robin interface condition for
  fluid-structure coupled simulations},\ }\href@noop {} {\bibfield  {journal}
  {\bibinfo  {journal} {International Journal for Numerical Methods in
  Engineering}\ }\textbf {\bibinfo {volume} {122}},\ \bibinfo {pages} {5176}
  (\bibinfo {year} {2021})}\BibitemShut {NoStop}%
\bibitem [{\citenamefont {Gurugubelli}\ and\ \citenamefont
  {Jaiman}(2019)}]{gurugubelli2019large}%
  \BibitemOpen
  \bibfield  {author} {\bibinfo {author} {\bibfnamefont {P.~S.}\ \bibnamefont
  {Gurugubelli}}\ and\ \bibinfo {author} {\bibfnamefont {R.~K.}\ \bibnamefont
  {Jaiman}},\ }\bibfield  {title} {\bibinfo {title} {Large amplitude flapping
  of an inverted elastic foil in uniform flow with spanwise periodicity},\
  }\href@noop {} {\bibfield  {journal} {\bibinfo  {journal} {Journal of Fluids
  and Structures}\ }\textbf {\bibinfo {volume} {90}},\ \bibinfo {pages} {139}
  (\bibinfo {year} {2019})}\BibitemShut {NoStop}%
\bibitem [{\citenamefont {Tang}\ \emph {et~al.}(2015)\citenamefont {Tang},
  \citenamefont {Liu},\ and\ \citenamefont {Lu}}]{tang2015dynamics}%
  \BibitemOpen
  \bibfield  {author} {\bibinfo {author} {\bibfnamefont {C.}~\bibnamefont
  {Tang}}, \bibinfo {author} {\bibfnamefont {N.-S.}\ \bibnamefont {Liu}},\ and\
  \bibinfo {author} {\bibfnamefont {X.-Y.}\ \bibnamefont {Lu}},\ }\bibfield
  {title} {\bibinfo {title} {Dynamics of an inverted flexible plate in a
  uniform flow},\ }\href@noop {} {\bibfield  {journal} {\bibinfo  {journal}
  {Physics of Fluids}\ }\textbf {\bibinfo {volume} {27}} (\bibinfo {year}
  {2015})}\BibitemShut {NoStop}%
\bibitem [{\citenamefont {Shoele}\ and\ \citenamefont
  {Mittal}(2016)}]{shoele2016energy}%
  \BibitemOpen
  \bibfield  {author} {\bibinfo {author} {\bibfnamefont {K.}~\bibnamefont
  {Shoele}}\ and\ \bibinfo {author} {\bibfnamefont {R.}~\bibnamefont
  {Mittal}},\ }\bibfield  {title} {\bibinfo {title} {Energy harvesting by
  flow-induced flutter in a simple model of an inverted piezoelectric flag},\
  }\href@noop {} {\bibfield  {journal} {\bibinfo  {journal} {Journal of Fluid
  Mechanics}\ }\textbf {\bibinfo {volume} {790}},\ \bibinfo {pages} {582}
  (\bibinfo {year} {2016})}\BibitemShut {NoStop}%
\bibitem [{\citenamefont {Goza}\ \emph {et~al.}(2018)\citenamefont {Goza},
  \citenamefont {Colonius},\ and\ \citenamefont {Sader}}]{goza2018global}%
  \BibitemOpen
  \bibfield  {author} {\bibinfo {author} {\bibfnamefont {A.}~\bibnamefont
  {Goza}}, \bibinfo {author} {\bibfnamefont {T.}~\bibnamefont {Colonius}},\
  and\ \bibinfo {author} {\bibfnamefont {J.~E.}\ \bibnamefont {Sader}},\
  }\bibfield  {title} {\bibinfo {title} {Global modes and nonlinear analysis of
  inverted-flag flapping},\ }\href@noop {} {\bibfield  {journal} {\bibinfo
  {journal} {Journal of Fluid Mechanics}\ }\textbf {\bibinfo {volume} {857}},\
  \bibinfo {pages} {312} (\bibinfo {year} {2018})}\BibitemShut {NoStop}%
\bibitem [{\citenamefont {Rinaldi}\ and\ \citenamefont
  {Pa{\"\i}doussis}(2012)}]{rinaldi2012theory}%
  \BibitemOpen
  \bibfield  {author} {\bibinfo {author} {\bibfnamefont {S.}~\bibnamefont
  {Rinaldi}}\ and\ \bibinfo {author} {\bibfnamefont {M.~P.}\ \bibnamefont
  {Pa{\"\i}doussis}},\ }\bibfield  {title} {\bibinfo {title} {Theory and
  experiments on the dynamics of a free-clamped cylinder in confined axial
  air-flow},\ }\href@noop {} {\bibfield  {journal} {\bibinfo  {journal}
  {Journal of Fluids and Structures}\ }\textbf {\bibinfo {volume} {28}},\
  \bibinfo {pages} {167} (\bibinfo {year} {2012})}\BibitemShut {NoStop}%
\bibitem [{\citenamefont {Abdelbaki}\ \emph {et~al.}(2018)\citenamefont
  {Abdelbaki}, \citenamefont {Pa{\"\i}doussis},\ and\ \citenamefont
  {Misra}}]{abdelbaki2018nonlinear}%
  \BibitemOpen
  \bibfield  {author} {\bibinfo {author} {\bibfnamefont {A.}~\bibnamefont
  {Abdelbaki}}, \bibinfo {author} {\bibfnamefont {M.~P.}\ \bibnamefont
  {Pa{\"\i}doussis}},\ and\ \bibinfo {author} {\bibfnamefont {A.}~\bibnamefont
  {Misra}},\ }\bibfield  {title} {\bibinfo {title} {A nonlinear model for a
  free-clamped cylinder subjected to confined axial flow},\ }\href@noop {}
  {\bibfield  {journal} {\bibinfo  {journal} {Journal of Fluids and
  Structures}\ }\textbf {\bibinfo {volume} {80}},\ \bibinfo {pages} {390}
  (\bibinfo {year} {2018})}\BibitemShut {NoStop}%
\bibitem [{\citenamefont {Sader}\ \emph
  {et~al.}(2016{\natexlab{b}})\citenamefont {Sader}, \citenamefont
  {Huertas-Cerdeira},\ and\ \citenamefont {Gharib}}]{sader2016stability}%
  \BibitemOpen
  \bibfield  {author} {\bibinfo {author} {\bibfnamefont {J.~E.}\ \bibnamefont
  {Sader}}, \bibinfo {author} {\bibfnamefont {C.}~\bibnamefont
  {Huertas-Cerdeira}},\ and\ \bibinfo {author} {\bibfnamefont {M.}~\bibnamefont
  {Gharib}},\ }\bibfield  {title} {\bibinfo {title} {Stability of slender
  inverted flags and rods in uniform steady flow},\ }\href@noop {} {\bibfield
  {journal} {\bibinfo  {journal} {Journal of Fluid Mechanics}\ }\textbf
  {\bibinfo {volume} {809}},\ \bibinfo {pages} {873} (\bibinfo {year}
  {2016}{\natexlab{b}})}\BibitemShut {NoStop}%
\bibitem [{\citenamefont {Tavallaeinejad}\ \emph {et~al.}(2018)\citenamefont
  {Tavallaeinejad}, \citenamefont {Pa{\"\i}doussis},\ and\ \citenamefont
  {Legrand}}]{tavallaeinejad2018nonlinear}%
  \BibitemOpen
  \bibfield  {author} {\bibinfo {author} {\bibfnamefont {M.}~\bibnamefont
  {Tavallaeinejad}}, \bibinfo {author} {\bibfnamefont {M.~P.}\ \bibnamefont
  {Pa{\"\i}doussis}},\ and\ \bibinfo {author} {\bibfnamefont {M.}~\bibnamefont
  {Legrand}},\ }\bibfield  {title} {\bibinfo {title} {Nonlinear static response
  of low-aspect-ratio inverted flags subjected to a steady flow},\ }\href@noop
  {} {\bibfield  {journal} {\bibinfo  {journal} {Journal of Fluids and
  Structures}\ }\textbf {\bibinfo {volume} {83}},\ \bibinfo {pages} {413}
  (\bibinfo {year} {2018})}\BibitemShut {NoStop}%
\bibitem [{\citenamefont {Khalak}\ and\ \citenamefont
  {Williamson}(1999)}]{khalak1999motions}%
  \BibitemOpen
  \bibfield  {author} {\bibinfo {author} {\bibfnamefont {A.}~\bibnamefont
  {Khalak}}\ and\ \bibinfo {author} {\bibfnamefont {C.~H.}\ \bibnamefont
  {Williamson}},\ }\bibfield  {title} {\bibinfo {title} {Motions, forces and
  mode transitions in vortex-induced vibrations at low mass-damping},\
  }\href@noop {} {\bibfield  {journal} {\bibinfo  {journal} {Journal of fluids
  and Structures}\ }\textbf {\bibinfo {volume} {13}},\ \bibinfo {pages} {813}
  (\bibinfo {year} {1999})}\BibitemShut {NoStop}%
\bibitem [{\citenamefont {Sarpkaya}(2004)}]{sarpkaya2004critical}%
  \BibitemOpen
  \bibfield  {author} {\bibinfo {author} {\bibfnamefont {T.}~\bibnamefont
  {Sarpkaya}},\ }\bibfield  {title} {\bibinfo {title} {A critical review of the
  intrinsic nature of vortex-induced vibrations},\ }\href@noop {} {\bibfield
  {journal} {\bibinfo  {journal} {Journal of fluids and structures}\ }\textbf
  {\bibinfo {volume} {19}},\ \bibinfo {pages} {389} (\bibinfo {year}
  {2004})}\BibitemShut {NoStop}%
\bibitem [{\citenamefont {Tavallaeinejad}\ \emph
  {et~al.}(2020{\natexlab{b}})\citenamefont {Tavallaeinejad}, \citenamefont
  {Legrand},\ and\ \citenamefont {Paidoussis}}]{tavallaeinejad2020nonlinear}%
  \BibitemOpen
  \bibfield  {author} {\bibinfo {author} {\bibfnamefont {M.}~\bibnamefont
  {Tavallaeinejad}}, \bibinfo {author} {\bibfnamefont {M.}~\bibnamefont
  {Legrand}},\ and\ \bibinfo {author} {\bibfnamefont {M.~P.}\ \bibnamefont
  {Paidoussis}},\ }\bibfield  {title} {\bibinfo {title} {Nonlinear dynamics of
  slender inverted flags in uniform steady flows},\ }\href@noop {} {\bibfield
  {journal} {\bibinfo  {journal} {Journal of Sound and Vibration}\ }\textbf
  {\bibinfo {volume} {467}},\ \bibinfo {pages} {115048} (\bibinfo {year}
  {2020}{\natexlab{b}})}\BibitemShut {NoStop}%
\bibitem [{\citenamefont {Sirovich}(1987)}]{sirovich1987turbulence}%
  \BibitemOpen
  \bibfield  {author} {\bibinfo {author} {\bibfnamefont {L.}~\bibnamefont
  {Sirovich}},\ }\bibfield  {title} {\bibinfo {title} {Turbulence and the
  dynamics of coherent structures, parts i, ii and iii},\ }\href@noop {}
  {\bibfield  {journal} {\bibinfo  {journal} {Quart. Appl. Math.}\ ,\ \bibinfo
  {pages} {561}} (\bibinfo {year} {1987})}\BibitemShut {NoStop}%
\bibitem [{\citenamefont {Aris}(2012)}]{aris2012vectors}%
  \BibitemOpen
  \bibfield  {author} {\bibinfo {author} {\bibfnamefont {R.}~\bibnamefont
  {Aris}},\ }\href@noop {} {\emph {\bibinfo {title} {Vectors, tensors and the
  basic equations of fluid mechanics}}}\ (\bibinfo  {publisher} {Courier
  Corporation},\ \bibinfo {year} {2012})\BibitemShut {NoStop}%
\bibitem [{\citenamefont {Ogden}(1997)}]{ogden1997non}%
  \BibitemOpen
  \bibfield  {author} {\bibinfo {author} {\bibfnamefont {R.~W.}\ \bibnamefont
  {Ogden}},\ }\href@noop {} {\emph {\bibinfo {title} {Non-linear elastic
  deformations}}}\ (\bibinfo  {publisher} {Courier Corporation},\ \bibinfo
  {year} {1997})\BibitemShut {NoStop}%
\bibitem [{\citenamefont {Helenbrook}(2003)}]{helenbrook2003mesh}%
  \BibitemOpen
  \bibfield  {author} {\bibinfo {author} {\bibfnamefont {B.~T.}\ \bibnamefont
  {Helenbrook}},\ }\bibfield  {title} {\bibinfo {title} {Mesh deformation using
  the biharmonic operator},\ }\href@noop {} {\bibfield  {journal} {\bibinfo
  {journal} {International journal for numerical methods in engineering}\
  }\textbf {\bibinfo {volume} {56}},\ \bibinfo {pages} {1007} (\bibinfo {year}
  {2003})}\BibitemShut {NoStop}%
\bibitem [{\citenamefont {Ghattas}\ and\ \citenamefont
  {Li}(1995)}]{ghattas1995variational}%
  \BibitemOpen
  \bibfield  {author} {\bibinfo {author} {\bibfnamefont {O.}~\bibnamefont
  {Ghattas}}\ and\ \bibinfo {author} {\bibfnamefont {X.}~\bibnamefont {Li}},\
  }\bibfield  {title} {\bibinfo {title} {A variational finite element method
  for stationary nonlinear fluid—solid interaction},\ }\href@noop {}
  {\bibfield  {journal} {\bibinfo  {journal} {Journal of Computational
  Physics}\ }\textbf {\bibinfo {volume} {121}},\ \bibinfo {pages} {347}
  (\bibinfo {year} {1995})}\BibitemShut {NoStop}%
\bibitem [{\citenamefont {Jiang}\ \emph
  {et~al.}(2022{\natexlab{a}})\citenamefont {Jiang}, \citenamefont {Luo},
  \citenamefont {Zhang}, \citenamefont {Wu},\ and\ \citenamefont
  {Yi}}]{jiang2022global}%
  \BibitemOpen
  \bibfield  {author} {\bibinfo {author} {\bibfnamefont {H.-K.}\ \bibnamefont
  {Jiang}}, \bibinfo {author} {\bibfnamefont {K.}~\bibnamefont {Luo}}, \bibinfo
  {author} {\bibfnamefont {Z.-Y.}\ \bibnamefont {Zhang}}, \bibinfo {author}
  {\bibfnamefont {J.}~\bibnamefont {Wu}},\ and\ \bibinfo {author}
  {\bibfnamefont {H.-L.}\ \bibnamefont {Yi}},\ }\bibfield  {title} {\bibinfo
  {title} {Global linear instability analysis of thermal convective flow using
  the linearized lattice boltzmann method},\ }\href@noop {} {\bibfield
  {journal} {\bibinfo  {journal} {Journal of Fluid Mechanics}\ }\textbf
  {\bibinfo {volume} {944}},\ \bibinfo {pages} {A31} (\bibinfo {year}
  {2022}{\natexlab{a}})}\BibitemShut {NoStop}%
\bibitem [{\citenamefont {Jiang}\ \emph
  {et~al.}(2022{\natexlab{b}})\citenamefont {Jiang}, \citenamefont {Zhang},
  \citenamefont {Zhang}, \citenamefont {Luo},\ and\ \citenamefont
  {Yi}}]{jiang2022instability}%
  \BibitemOpen
  \bibfield  {author} {\bibinfo {author} {\bibfnamefont {H.-K.}\ \bibnamefont
  {Jiang}}, \bibinfo {author} {\bibfnamefont {Y.}~\bibnamefont {Zhang}},
  \bibinfo {author} {\bibfnamefont {Z.-Y.}\ \bibnamefont {Zhang}}, \bibinfo
  {author} {\bibfnamefont {K.}~\bibnamefont {Luo}},\ and\ \bibinfo {author}
  {\bibfnamefont {H.-L.}\ \bibnamefont {Yi}},\ }\bibfield  {title} {\bibinfo
  {title} {Instability and bifurcations of electro-thermo-convection in a
  tilted square cavity filled with dielectric liquid},\ }\href@noop {}
  {\bibfield  {journal} {\bibinfo  {journal} {Physics of Fluids}\ }\textbf
  {\bibinfo {volume} {34}} (\bibinfo {year} {2022}{\natexlab{b}})}\BibitemShut
  {NoStop}%
\bibitem [{\citenamefont {Lehoucq}\ \emph {et~al.}(1998)\citenamefont
  {Lehoucq}, \citenamefont {Sorensen},\ and\ \citenamefont
  {Yang}}]{lehoucq1998arpack}%
  \BibitemOpen
  \bibfield  {author} {\bibinfo {author} {\bibfnamefont {R.~B.}\ \bibnamefont
  {Lehoucq}}, \bibinfo {author} {\bibfnamefont {D.~C.}\ \bibnamefont
  {Sorensen}},\ and\ \bibinfo {author} {\bibfnamefont {C.}~\bibnamefont
  {Yang}},\ }\href@noop {} {\emph {\bibinfo {title} {ARPACK users' guide:
  solution of large-scale eigenvalue problems with implicitly restarted Arnoldi
  methods}}}\ (\bibinfo  {publisher} {SIAM},\ \bibinfo {year}
  {1998})\BibitemShut {NoStop}%
\bibitem [{\citenamefont {Eckart}\ and\ \citenamefont
  {Young}(1936)}]{eckart1936approximation}%
  \BibitemOpen
  \bibfield  {author} {\bibinfo {author} {\bibfnamefont {C.}~\bibnamefont
  {Eckart}}\ and\ \bibinfo {author} {\bibfnamefont {G.}~\bibnamefont {Young}},\
  }\bibfield  {title} {\bibinfo {title} {The approximation of one matrix by
  another of lower rank},\ }\href@noop {} {\bibfield  {journal} {\bibinfo
  {journal} {Psychometrika}\ }\textbf {\bibinfo {volume} {1}},\ \bibinfo
  {pages} {211} (\bibinfo {year} {1936})}\BibitemShut {NoStop}%
\bibitem [{\citenamefont {Guckenheimer}\ and\ \citenamefont
  {Holmes}(2013)}]{guckenheimer2013nonlinear}%
  \BibitemOpen
  \bibfield  {author} {\bibinfo {author} {\bibfnamefont {J.}~\bibnamefont
  {Guckenheimer}}\ and\ \bibinfo {author} {\bibfnamefont {P.}~\bibnamefont
  {Holmes}},\ }\href@noop {} {\emph {\bibinfo {title} {Nonlinear oscillations,
  dynamical systems, and bifurcations of vector fields}}},\ Vol.~\bibinfo
  {volume} {42}\ (\bibinfo  {publisher} {Springer Science \& Business Media},\
  \bibinfo {year} {2013})\BibitemShut {NoStop}%
\bibitem [{\citenamefont {Aranson}\ and\ \citenamefont
  {Kramer}(2002)}]{aranson2002world}%
  \BibitemOpen
  \bibfield  {author} {\bibinfo {author} {\bibfnamefont {I.~S.}\ \bibnamefont
  {Aranson}}\ and\ \bibinfo {author} {\bibfnamefont {L.}~\bibnamefont
  {Kramer}},\ }\bibfield  {title} {\bibinfo {title} {The world of the complex
  ginzburg-landau equation},\ }\href@noop {} {\bibfield  {journal} {\bibinfo
  {journal} {Reviews of modern physics}\ }\textbf {\bibinfo {volume} {74}},\
  \bibinfo {pages} {99} (\bibinfo {year} {2002})}\BibitemShut {NoStop}%
\bibitem [{\citenamefont {Kang}\ \emph {et~al.}(2017)\citenamefont {Kang},
  \citenamefont {Meyer}, \citenamefont {Mutabazi},\ and\ \citenamefont
  {Yoshikawa}}]{kang2017radial}%
  \BibitemOpen
  \bibfield  {author} {\bibinfo {author} {\bibfnamefont {C.}~\bibnamefont
  {Kang}}, \bibinfo {author} {\bibfnamefont {A.}~\bibnamefont {Meyer}},
  \bibinfo {author} {\bibfnamefont {I.}~\bibnamefont {Mutabazi}},\ and\
  \bibinfo {author} {\bibfnamefont {H.~N.}\ \bibnamefont {Yoshikawa}},\
  }\bibfield  {title} {\bibinfo {title} {Radial buoyancy effects on momentum
  and heat transfer in a circular couette flow},\ }\href@noop {} {\bibfield
  {journal} {\bibinfo  {journal} {Physical Review Fluids}\ }\textbf {\bibinfo
  {volume} {2}},\ \bibinfo {pages} {053901} (\bibinfo {year}
  {2017})}\BibitemShut {NoStop}%
\bibitem [{\citenamefont {Kang}\ \emph {et~al.}(2019)\citenamefont {Kang},
  \citenamefont {Meyer}, \citenamefont {Yoshikawa},\ and\ \citenamefont
  {Mutabazi}}]{kang2019numerical}%
  \BibitemOpen
  \bibfield  {author} {\bibinfo {author} {\bibfnamefont {C.}~\bibnamefont
  {Kang}}, \bibinfo {author} {\bibfnamefont {A.}~\bibnamefont {Meyer}},
  \bibinfo {author} {\bibfnamefont {H.~N.}\ \bibnamefont {Yoshikawa}},\ and\
  \bibinfo {author} {\bibfnamefont {I.}~\bibnamefont {Mutabazi}},\ }\bibfield
  {title} {\bibinfo {title} {Numerical study of thermal convection induced by
  centrifugal buoyancy in a rotating cylindrical annulus},\ }\href@noop {}
  {\bibfield  {journal} {\bibinfo  {journal} {Physical Review Fluids}\ }\textbf
  {\bibinfo {volume} {4}},\ \bibinfo {pages} {043501} (\bibinfo {year}
  {2019})}\BibitemShut {NoStop}%
\bibitem [{\citenamefont {Golub}\ and\ \citenamefont
  {Van~Loan}(2013)}]{golub2013matrix}%
  \BibitemOpen
  \bibfield  {author} {\bibinfo {author} {\bibfnamefont {G.~H.}\ \bibnamefont
  {Golub}}\ and\ \bibinfo {author} {\bibfnamefont {C.~F.}\ \bibnamefont
  {Van~Loan}},\ }\href@noop {} {\emph {\bibinfo {title} {Matrix
  computations}}}\ (\bibinfo  {publisher} {JHU press},\ \bibinfo {year}
  {2013})\BibitemShut {NoStop}%
\bibitem [{\citenamefont {Gurugubelli}\ and\ \citenamefont
  {Jaiman}(2015)}]{gurugubelli2015self}%
  \BibitemOpen
  \bibfield  {author} {\bibinfo {author} {\bibfnamefont {P.~S.}\ \bibnamefont
  {Gurugubelli}}\ and\ \bibinfo {author} {\bibfnamefont {R.~K.}\ \bibnamefont
  {Jaiman}},\ }\bibfield  {title} {\bibinfo {title} {Self-induced flapping
  dynamics of a flexible inverted foil in a uniform flow},\ }\href@noop {}
  {\bibfield  {journal} {\bibinfo  {journal} {Journal of Fluid Mechanics}\
  }\textbf {\bibinfo {volume} {781}},\ \bibinfo {pages} {657} (\bibinfo {year}
  {2015})}\BibitemShut {NoStop}%
\bibitem [{\citenamefont {Wick}(2012)}]{wick2012stability}%
  \BibitemOpen
  \bibfield  {author} {\bibinfo {author} {\bibfnamefont {T.}~\bibnamefont
  {Wick}},\ }\bibfield  {title} {\bibinfo {title} {Stability estimates and
  numerical comparison of second order time-stepping schemes for
  fluid-structure interactions},\ }in\ \href@noop {} {\emph {\bibinfo
  {booktitle} {Numerical Mathematics and Advanced Applications 2011:
  Proceedings of ENUMATH 2011, the 9th European Conference on Numerical
  Mathematics and Advanced Applications, Leicester, September 2011}}}\
  (\bibinfo {organization} {Springer},\ \bibinfo {year} {2012})\ pp.\ \bibinfo
  {pages} {625--632}\BibitemShut {NoStop}%
\bibitem [{\citenamefont {Aln{\ae}s}\ \emph {et~al.}(2015)\citenamefont
  {Aln{\ae}s}, \citenamefont {Blechta}, \citenamefont {Hake}, \citenamefont
  {Johansson}, \citenamefont {Kehlet}, \citenamefont {Logg}, \citenamefont
  {Richardson}, \citenamefont {Ring}, \citenamefont {Rognes},\ and\
  \citenamefont {Wells}}]{alnaes2015fenics}%
  \BibitemOpen
  \bibfield  {author} {\bibinfo {author} {\bibfnamefont {M.}~\bibnamefont
  {Aln{\ae}s}}, \bibinfo {author} {\bibfnamefont {J.}~\bibnamefont {Blechta}},
  \bibinfo {author} {\bibfnamefont {J.}~\bibnamefont {Hake}}, \bibinfo {author}
  {\bibfnamefont {A.}~\bibnamefont {Johansson}}, \bibinfo {author}
  {\bibfnamefont {B.}~\bibnamefont {Kehlet}}, \bibinfo {author} {\bibfnamefont
  {A.}~\bibnamefont {Logg}}, \bibinfo {author} {\bibfnamefont {C.}~\bibnamefont
  {Richardson}}, \bibinfo {author} {\bibfnamefont {J.}~\bibnamefont {Ring}},
  \bibinfo {author} {\bibfnamefont {M.~E.}\ \bibnamefont {Rognes}},\ and\
  \bibinfo {author} {\bibfnamefont {G.~N.}\ \bibnamefont {Wells}},\ }\bibfield
  {title} {\bibinfo {title} {The fenics project version 1.5},\ }\href@noop {}
  {\bibfield  {journal} {\bibinfo  {journal} {Archive of numerical software}\
  }\textbf {\bibinfo {volume} {3}} (\bibinfo {year} {2015})}\BibitemShut
  {NoStop}%
\end{thebibliography}%

\end{document}